\documentclass[a4paper,fleqn]{cas-dc-mod}

\usepackage[authoryear]{natbib}

\usepackage{caption}
\usepackage{subcaption}
\usepackage{float}
\usepackage{placeins}
\usepackage{flafter}

\usepackage{url}

\begin{document}
\let\WriteBookmarks\relax
\def\floatpagepagefraction{1}
\def\textpagefraction{.001}

\shorttitle{Regular Asymptomatic Testing for COVID-19}

\shortauthors{}

\title[mode = title]{The role of regular asymptomatic testing in reducing the impact of a COVID-19 wave}

\author[1]{Miguel E. P. Silva}

\cormark[1]


\ead{miguel.silva@manchester.ac.uk}

\affiliation[1]{organization={Department of Computer Science, University of Manchester},
    country={United Kingdom}}

\author[2,3]{Martyn Fyles}

\affiliation[2]{organization={Department of Mathematics, University of Manchester},
    country={United Kingdom}}
\affiliation[3]{organization={The Alan Turing Institute, London},
    country={United Kingdom}}

\author[4]{Li Pi}

\affiliation[4]{organization={Big Data Institute, Li Ka Shing Centre for Health Information and Discovery, University of Oxford},country={United Kingdom}}

\author[4,5,6]{Jasmina Panovska-Griffiths}

\affiliation[5]{organization={The Queen's College, University of Oxford},
    country={United Kingdom}}

\affiliation[6]{organization={Pandemic Sciences Institute, Nuffield Department of Medicine, University of Oxford},
    country={United Kingdom}}

\author[2]{Thomas House}

\author[1]{Caroline Jay}

\author[7,8]{Elizabeth Fearon}

\affiliation[7]{organization={Department of Global Health and Development, Faculty of Public Health and Policy, London School of Hygiene and Tropical Medicine},country={United Kingdom}}

\affiliation[8]{organization={Centre for Mathematical Modelling of Infectious Diseases, London School of Hygiene and Tropical Medicine},country={United Kingdom}}

\cortext[cor1]{Corresponding author}

\begin{abstract}
Testing for infection with SARS-CoV-2 is an important intervention in reducing
onwards transmission of COVID-19, particularly when combined with the isolation and contact-tracing of positive cases. Many countries with
the capacity to do so have made use of lab-processed Polymerase Chain Reaction (PCR) testing
targeted at individuals with symptoms and the contacts of confirmed cases. Alternatively, Lateral Flow Tests (LFTs) are able to deliver a result quickly, without lab-processing and at a relatively low cost. Their adoption can support regular mass
asymptomatic testing, allowing earlier detection of infection and isolation of
infectious individuals. In this paper we extend and apply the
agent-based epidemic modelling framework \textit{Covasim} to explore the impact
of regular asymptomatic testing on the peak and total number of infections
in an emerging COVID-19 wave. We explore testing with LFTs at different
frequency levels within a population with high levels of immunity and with
background symptomatic PCR testing, case isolation and contact tracing for testing. The effectiveness of
regular asymptomatic testing was compared with `lockdown' interventions seeking
to reduce the number of non-household contacts across the whole population through
measures such as mandating working from home and restrictions on gatherings.
Since regular asymptomatic testing
 requires only those with a positive result to reduce contact, while lockdown measures require the whole population to reduce
contact, any policy decision that seeks to trade off harms from infection
against other harms will not automatically favour one over the other. Our
results demonstrate that, where such a trade off is being made, at moderate rates
of early exponential growth regular asymptomatic testing has the potential to achieve significant infection control without the wider harms associated with additional lockdown measures.
\end{abstract}


\maketitle

\section{Introduction}

Severe acute respiratory syndrome coronavirus 2 (SARS-CoV-2), the virus causing COVID-19, has continued to spread in the United Kingdom (UK) throughout 2020, 2021 and 2022 facilitated by the emergence and progressive dominance of new viral lineages such as the Alpha, Delta and Omicron variants of concern \citep{WHO_variants}. At time of writing, over 22 million confirmed cases and over 175 thousand deaths related to COVID-19 have been reported in the UK \citep{coviddashboard}.

Each wave of infection has posed questions about the appropriate interventions
to be made to mitigate the impacts of the disease. While the first two waves
of infection in the UK were mainly managed through `lockdown' measures
involving policies such as mandated working from home and limits on
out-of-household mixing, by 2021 vaccines had become available, with over 30
million first doses distributed by the end of March 2021. At time of writing,
over 53 million first, over 50 million second, and over 39 million third doses
have been distributed in the UK \citep{coviddashboard}.

As 2021 progressed, and even
prior to the arrival of the Delta and Omicron variants, there was a concern
that a fast growth in cases -- particularly driven by a variant either evading
immunity or with enhanced inherent transmissibility -- might lead to large
absolute numbers of severe cases, putting pressure on an already stretched
healthcare system, and leading to a large number of deaths
\citep{dyson_possible_2021}.  The major legal restrictions to economic,
educational and social activities that had been implemented as emergency
measures to disrupt transmission during the previous 18
months, which were believed to be of increasing harm \citep{Viner_2022}
and decreasing efficacy over time \citep{Delussu_2022}, provided a strong
motivation in late 2021 to find ways of reducing SARS-CoV-2 transmission using
other methods.

An alternative targeted approach to ongoing lockdowns and restrictions on social
gatherings, is seeking to identify those who are infectious and restricting isolation
to only those individuals. This is the goal of Test-Trace-Isolate (TTI) strategies, but
using TTI alone to control growth was found to be challenging early in the
pandemic \citep{Ferretti_2020, Davis_2021, Fyles_2021}. Identifying individuals who are infectious before they have the opportunity to make contacts is difficult because many cases are only mildly symptomatic or asymptomatic, and presymptomatic transmission is common, making symptom-based case ascertainment unreliable. In the UK, from the start of the pandemic, the government invested heavily in Polymerase Chain Reaction (PCR) testing infrastructure, which was primarily used in the community to detect infections among individuals experiencing one of a set of symptoms. PCR testing involves taking samples with swabs (of the naso-pharynx, as advised by test providers in the UK), and transporting the samples to one of a network of laboratories where the sample is amplified to detect any fragments of SARS-CoV-2 RNA. Later in the pandemic life cycle, an alternative technology provided by lateral flow tests (LFTs) -- alternatively called rapid antigen tests (RATs) or lateral flow device (LFD) tests -- was widely deployed in the UK. These tests are inexpensive cartridges that deliver results within 15-30 minutes, rather than in 1-2 days for PCR tests, and do not require laboratories with their associated staff or logistic networks.

While it is possible to deploy PCR at scale -- as has been done most
prominently in China \citep{li_comprehensive_2021} -- the speed and low cost of
LFTs makes these more attractive for large population testing programmes.
Further, \citet{Larremore:2021} argue that the key criterion for a test being
useful for these purposes is sensitivity during the infectious period rather
than sensitivity to any level of detectable virus. According to this criterion,
LFTs are not expected to be substantially inferior to PCR
\citep{ke_daily_2022}. In high prevalence settings ($\geq$ 5\%), the World Health Organisation (WHO) recommends prioritisation of LFTs as they can be quickly produced in large
quantities and the results can be obtained rapidly on site and without the need
for a laboratory \citep{WHO_LFTs}. A positive LFT result will then allow
immediate isolation of suspected cases and timely contact tracing.

By providing a rapid result and reducing the need for resource-intensive laboratory and transport infrastructure subject to capacity limitations, LFTs offered the opportunity to expand case finding strategies to asymptomatic as well as symptomatic individuals on a wider community level. Mass use of single LFDs targeted to asymptomatic individuals or regardless of symptoms had a significant impact when deployed nationally in Slovakia \citep{Slovakia_Mass_Test} and locally in Liverpool \citep{Liverpool_SMART}, although these schemes also required significant human resources in their delivery. Through the first half of 2021, the UK applied setting-specific asymptomatic testing programmes, such as in care homes, healthcare settings, schools and workplaces and some evaluation of the potential efficacy and effectiveness of these programmes has occurred \citep{YOUNG20211217, Buckle_2021, Leng_2022}.

At the start of Autumn 2021, in order to avoid both the direct harms of a large Winter epidemic wave and implications for healthcare capacity, as well as the indirect harms of restrictive lockdown measures and social contact restrictions, there was a need to assess the potential effectiveness of a regular population-wide asymptomatic screening programme using LFD tests. Our work aims to explore variations on such a policy, against the background of the existing symptomatic testing and tracing programme, using an adaptation of a published agent-based model called \textit{Covasim} \citep{kerr2021covasim}.

Specifically, we answer the following five questions:
\begin{enumerate}
\item What is the effectiveness of regular LFD testing among asymptomatic
individuals in reducing the total number of infections and the size of the
peak:
\begin{itemize}
\item According to testing frequency?
\item According to the level of testing and isolation uptake?
\end{itemize}
\item What percentage of non-household contact reduction is required to reduce the number of infections to the same extent as a
regular asymptomatic testing policy?
\item How does adherence to the following TTI strategies affect number of infections?:
\begin{itemize}
    \item Isolation on symptom-onset;
    \item Testing of symptomatic individuals;
    \item Testing after contact tracing of an identified positive case.
\end{itemize}
\item For the range of testing policies, what are the associated ‘costs’ in
terms of numbers of days people are asked to isolate at home and numbers of
tests required?
\item What are the benefits of testing with a single LFD test compared to
a single PCR test when symptomatic?

\end{enumerate}

\section{Methods}

\subsection{Model Overview}

We use Covasim, a previously described individual-based model \citep{kerr2021covasim}, which has been used in the UK to evaluate the impact of different TTI strategies when schools reopened after the first national lockdown \citep{jpg2020tti}, to explore the impact of wearing face coverings in schools \citep{jpg2021masks}, to simulate different lockdown lifting strategies \citep{jpg2021step1, jpgroadmap2022}, and as part of the UK Health Security Agency's epidemic nowcasting \citep{ukhsaR}.

This study uses a modified version of Covasim based
on release 3.0.2 to consider different asymptomatic testing regimes and choices of testing technology for symptomatic testing, across a range of adherence levels, data-informed social contact patterns and epidemic growth rates. The modified Covasim and the code for running all simulations
contained in this paper are available at
\url{https://github.com/TTI-modelling/covasim} and
\url{https://github.com/migueleps/RegularAsymptomaticTesting}.

Treating each individual as an agent, \textit{Covasim}
incorporates individual transmission variability, age-specific disease
progression probabilities, immunity dynamics, and the effect of various
vaccination and non-pharmaceutical interventions (NPIs). Each timestep in the
model represents one day and is split into three main phases: interventions,
immunity and transmission. Interventions include both non-pharmaceutical
interventions, like contact tracing, and pharmaceutical ones, like vaccination.
Immunity in the model is implemented through neutralizing antibody
trajectories, which are mapped to reductions in transmissibility, susceptibility
and disease severity, a process described in~\cite{COVASIM_IMMUNITY}.
Transmission is calculated on a per connection basis, controlled by a parameter
$\beta$ representing the transmission probability. This parameter is affected
by the layer the contact occurred in, viral load and other factors, such as age-based
susceptibility. In
our analysis, we used the majority of default parameters regarding transmission
and disease progression in \emph{Covasim}, while the parameters specific to our
simulations and studied interventions are documented in
Appendix~\ref{app:parameters}.

Infectiousness of an individual is expected to be directly dependent on their
viral load. However, it is difficult to obtain data on the distribution of
viral load trajectories in general populations and the relationship between
viral load and infectiousness is not known. As such, we find it necessary to
take the simplifying assumption that the infectiousness of agents depends
solely on their infectious age. We base our model of infectiousness upon the
shape of the generation time distribution, defined as the distribution of times
from infection to transmission, estimated by Ferretti et al.
\citep{Ferretti2020.09.04.20188516} who combined several different datasets of
infector-contact pairs to produce an estimate for the generation time
distribution. Further, we calibrate the transmission control parameter $\beta$,
which controls overall infectiousness to produce epidemics with desired initial
growth rates.

\subsection{Contact Networks}

We simulated a population representative of a UK town, which consists of
100,000 individual agents linked by a 4-layered contact network (household,
school, work, and community). The contact structure (i.e. the average number of
contacts per person) and the transmission probability per contact are different
in each layer. In the household layer, the average number of contacts per
person is drawn from a Poisson distribution with mean of 2.25, informed by the
mean household size in the United Kingdom~\citep{UN_HH}. All members of a
household are fully connected to the remaining members of that household and
disconnected in this layer from any person outside the household.

We parameterized the structure of the remaining 3 contact layers using
age-dependent contact matrices obtained from the social contact surveys
CoMix~\citep{COMIX} and POLYMOD~\citep{POLYMOD}. The UK CoMix survey has been recording the dynamics
of individual’s  contact behaviours
bi-weekly in real time since March 2020, while POLYMOD was a comprehensive contact survey conducted in European countries in 2008, and therefore
represents pre-pandemic contact patterns. POLYMOD contact matrices for the UK
are available publicly online and we extracted contact matrices from
Rounds 22 to 27 (late August 2020 to September 2020) of the CoMix survey as they
documented the highest average number of contacts since the start of COVID-19
pandemic, up to the point the surveys were discontinued \citep{jarvis2021comix61}.

The CoMix surveys were designed to be consistent with POLYMOD, recording the age of
the participant, age of contact and the setting the contact occurred in. The
ages of participants and individual contacts are split into 9 age bins: 0 to 4
years, 5 to 11 years, 12 to 17, 18 to 29, 30 to 39, 40 to 49, 50 to 59, 60 to
69 and 70+ years. The CoMix survey asks participants to make a distinction
between individual contacts and ``group contacts"~\citep{gimma2022changes}, a
category for participants who are unable to recall details for all individual
contacts (for example, those working in public facing jobs). This category is
split into less fine-grained age divisions: 0 to 17, 18 to 64 and 65+ years.
These age bins are matched to the more detailed ones by sampling uniformly at
random an age bin in the appropriate range, for instance, to assign a contact
from the 0 to 17 age bin, we sample uniformly at random from the 0 to 4, 5 to
11 and 12 to 17 age bins. Applying the same analysis methodology as in the weekly CoMix reports,
we truncate the maximum number of contacts by each individual to 50 per day.

Using these age-mixing contact matrices, we generate each contact network layer
using a generalised configuration model in which the number of contacts of each
individual in each age bin is drawn from a negative binomial distribution and
then these age-specific desired contacts are paired uniformly at random.

Unlike the household, school and work layers, the community layer uses
Covasim's dynamic layer setting~\citep{kerr2021covasim}, wherein contacts are
regenerated after each timestep, while preserving the number of edges between
age bins.

\subsection{Interventions}

\begin{figure*}
    \centering

    \includegraphics[width=0.95\textwidth]{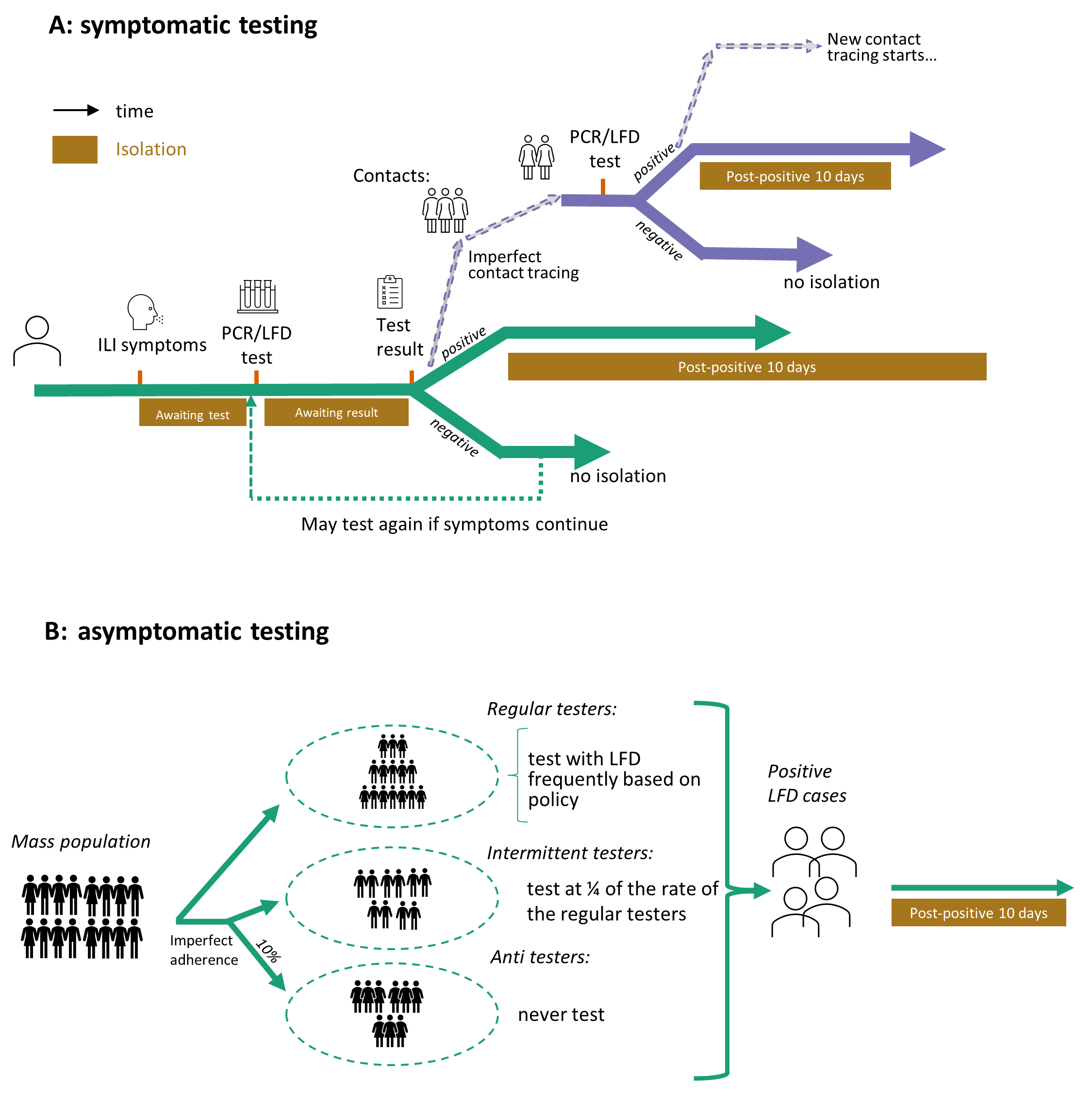}
    \caption{Diagram summarizing the workflow of symptomatic and asymptomatic testing strategies. }
    \label{fig:method}
\end{figure*}

We modeled government policy in our simulation setup through the inclusion of
four interventions: vaccination; isolation on symptom-onset; testing; and contact
tracing. Figure~\ref{fig:method} provides a summary of the relationships
between these interventions.

\subsubsection{Vaccination}

We assume a vaccination coverage scenario where 90\% of people aged over 40 have been vaccinated with 2 doses of the Astra Zeneca vaccine and 45\% of people aged under 40 have been vaccinated with 2 doses of the Pfizer vaccine. This leads to approximately 67\% of the population being vaccinated, which at the time we started this work, was considered a high vaccination coverage scenario, but it is likely to be underestimating population coverage at the time of writing. We also assume that the whole population is vaccinated from day 0, so the decaying vaccine efficacy resulting from decaying neutralizing antibodies is the same for all age groups. Through previous simulations, we calculated that this vaccination scenario reduces the total number of  infections by 70\% in an epidemic with growth rate 0.15, equivalent to a doubling time of 5 days.

\subsubsection{Isolation on symptom-onset}

Until April 2021, UK government policy advised that, upon developing at least one of three COVID symptoms (fever, new continuous cough or loss of smell or taste), people should self-isolate and request a PCR test, remaining in isolation until either receiving a negative result or for a period following a positive test (which varied, but was 10 days for the majority of the pandemic during 2020-2021). However, there was variation among the population in the ability to identify COVID symptoms correctly or to isolate prior to receiving test results. To take this into account, we implement isolation on symptom-onset with an adherence parameter, the probability that someone will preemptively isolate when they have developed symptoms.

As a respiratory disease, COVID shares symptoms with influenza-like illnesses (ILI), which can lead to people mistaking an ILI infection with COVID and preemptively self-isolating while waiting for a test result, despite not being infected with COVID. We model this scenario by picking randomly 1\% of the population to be infected with ILI symptoms every day and self-isolate with the same adherence probability.

Both groups of people isolating on symptom-onset are released from isolation if they test negative at least one day after developing symptoms.

\subsubsection{Testing}

Testing interventions allow the identification of infected individuals within a simulation. A diagnosed individual is put in isolation, which is implemented as a reduction to $\beta$ by 90\% in non-household contact layers. This isolation is always mandatory, regardless of the isolation on symptoms policy. Testing also unlocks contact tracing, which is implemented as a separate intervention but requires testing interventions to work.

We model testing differently depending on symptomatic status. In the symptomatic case, people with symptoms for COVID or ILI have a fixed probability of testing every day. Those who test positive are placed in at-home isolation for 10 days, regardless of their isolation status prior to the test.

In the asymptomatic case, we split the population into three groups: regular testers, intermittent testers and anti testers. Regular testers follow government advice rigorously, testing at periodic intervals. The rest of the population is split into 10\% anti testers, who never test if they are asymptomatic, and 90\% intermittent testers, who test at a quarter of the rate of the regular testers. This means that if 60\% of the population is testing regularly twice per week, then 6\% of the population never test if they are asymptomatic and 34\% of the population will test on average once every two weeks (7\% probability of testing per day).

As infection age at the point of testing is known to impact the effectiveness of TTI interventions, we added an extension to \emph{Covasim} which incorporates time dependent individual LFD and PCR test sensitivity trajectories. For these test sensitivity profiles, which describe the probability of infection being detected by a given test depending on infectious age, we use estimates obtained by \cite{hellewell2021estimating} using self-reported testing data from healthcare workers. As posterior samples were available for the test sensitivity profile parameters and there was significant uncertainty in the parameter estimates, we resample from the posterior before performing each simulation to ensure that the uncertainty in these estimates is fully captured in our model. We also include a testing delay for PCR tests, given by a Poisson distribution with mean 1.2, a swab error rate of 10\% (tests that return negative regardless of infection status) and a LFD test specificity of 99.7\% \citep{wolf2021lateral}.

\subsubsection{Contact tracing}

A positive test through a testing intervention triggers contact tracing. A proportion of contacts of the diagnosed individual are notified as contacts. The proportion of contacts and the time it takes to find them depends on the contact layer that the contact occurred in: in the household layer, we assume that all contacts are found within the same day of diagnosis; in the school and work layers, we assume that 45\% of contacts are found the day after the diagnosis; finally, in the community layer, we assume that only 15\% of contacts are found, taking 2 days from the case’s test result to do so.

We modified the default contact tracing interventions implemented in Covasim to better reflect UK policy during Autumn 2021. Instead of asking contacts of a diagnosed positive case to quarantine, they are asked to take a PCR test, which they do with probability controlled by an uptake parameter.

\subsection{Outcomes}

We focus the analysis of our model on the following outcomes:
\begin{itemize}
\item \textbf{Number of infections -} Total number of people who transition from a susceptible state to exposed, including reinfections.
\item \textbf{Height of new infections peak -} Maximum number of new infections in a single day.
\item \textbf{Number of tests -} Average number of tests, either PCR or LFD, that each individual in the simulation performs during the 180 days of simulation.
\item \textbf{Person days of isolation -} Average number of days that each individual is in isolation during the 180 days of simulation.
\end{itemize}

\section{Results}

We split our experiments into three sets of simulations, designed to answer the
five research questions we proposed. Each set of simulations was run with two
contact patterns, one obtained from POLYMOD and the other encompassing the
CoMix survey rounds 22 to 27, referring to late August 2020 and September 2020,
the time period when the number of average contacts was highest from the
beginning of the pandemic until November 2021, when the simulations were
executed.

We also calibrate the $\beta$ parameter per infectious contact to multiple
growth rates, for each of the contact patterns. The growth rates we use are
0.025, 0.05, 0.1, 0.15, 0.225 and 0.3, which correspond to approximate doubling
times of 1 month, 2 weeks, 1 week, 5 days, 3.5 days and 2.5 days, respectively.

\subsection{Asymptomatic testing and reducing number of contacts}
\label{sec:asympt}

To study the impact of increasing asymptomatic testing, we vary the percentage
of the population in the regular testing cohort from 0\% to 100\%, in intervals
of 10\%. We also vary the frequency of testing, with the regular groups testing
once per week, twice per week or every two days. The intermittent testers have
a daily probability of testing of 1/28, 1/14 and 1/7, respectively, meaning
that they will test on average once a month, once every two weeks or once per
week. In this set of simulations, we fix the adherence to isolation on
symptom-onset, the daily probability of testing if symptomatic and the proportion of
contacts that take up a PCR test to 40\%.

Government policy such as social distancing and sector closure (`lockdowns')
were introduced to prevent transmission of COVID by reducing the number of
contacts each person makes. To represent these interventions in Covasim, we
randomly remove a percentage of edges in the work, school and community layers
of the contact network (i.e.\ not the household network). We vary the
percentage of edges we remove from the network in increments of 5\%, from 0\%
to 100\%. For the simulations in this scenario, we remove all asymptomatic
testing and set adherence to isolation on symptom-onset, daily probability of
testing if symptomatic and the proportion of contacts that take up a PCR all to
the same value as before (40\%), to compare the impact of reducing
contacts with increasing asymptomatic testing.

\begin{figure*}[ht]
    \centering
    \begin{subfigure}{0.5\textwidth}
        \centering
        \includegraphics[width=\textwidth]{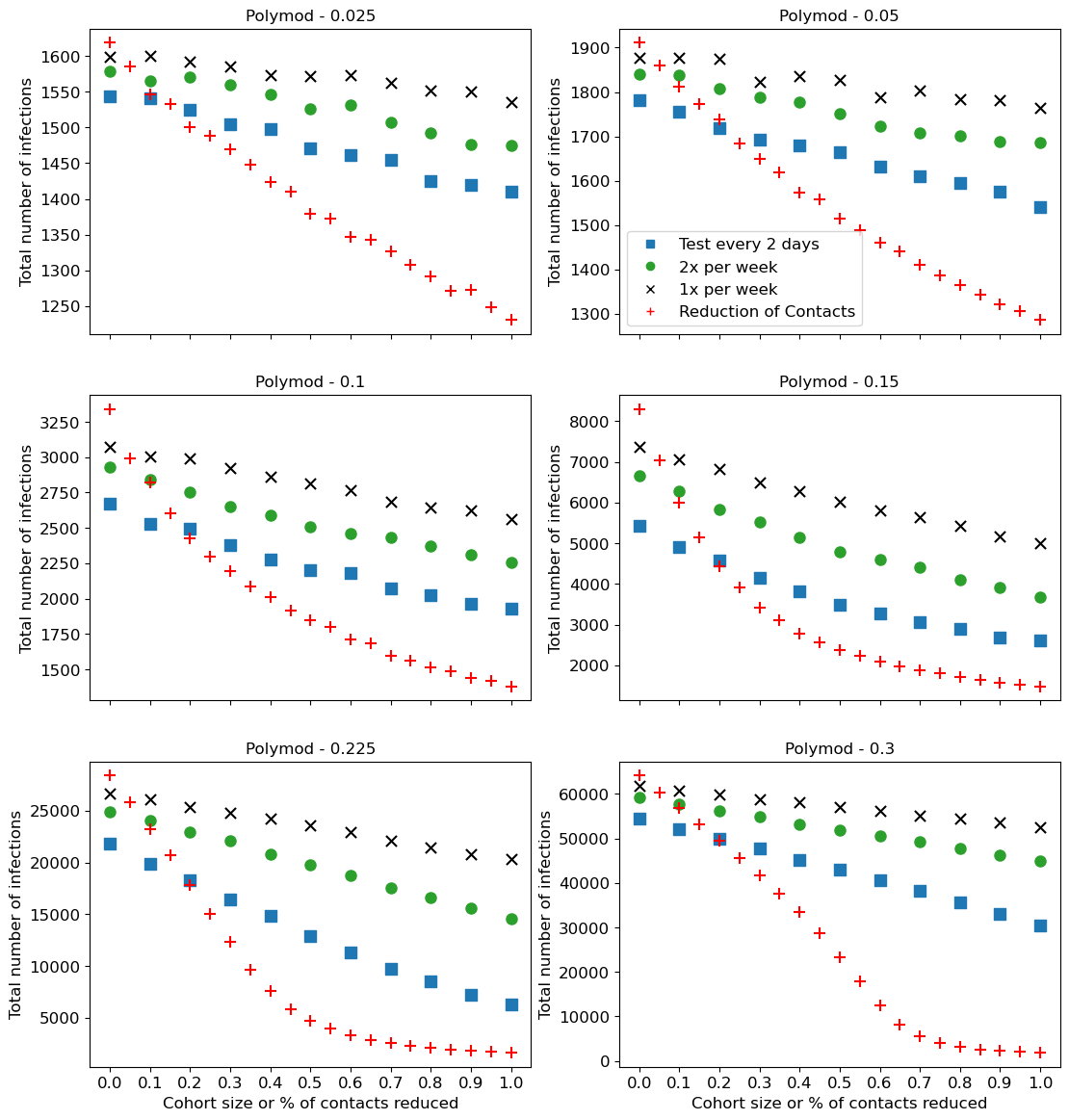}
        \caption{Contact patterns obtained from POLYMOD}
        \label{fig:asym_cases_polymod}
    \end{subfigure}%
    \begin{subfigure}{.5\textwidth}
        \centering
        \includegraphics[width=\textwidth]{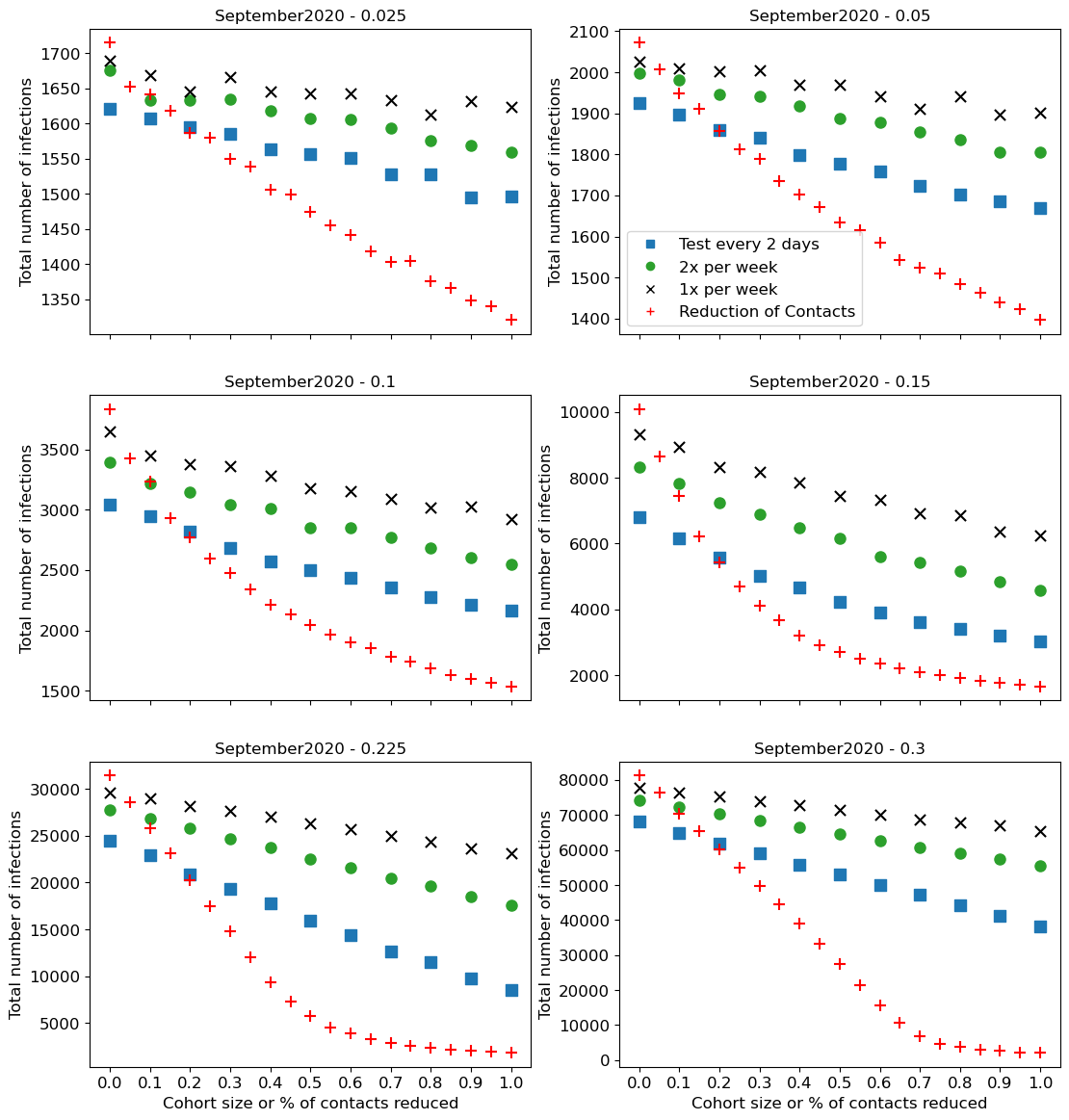}
        \caption{Contact patterns obtained from CoMix round 22 to 27 (late August 2020 to September 2020)}
        \label{fig:asym_cases_comix}
    \end{subfigure}
    \caption{Comparison of asymptomatic testing strategies against reduction of contacts for different growth rates and contact patterns. The x-axis represents the size of the regular tester cohort for the simulations with asymptomatic testing or the percentage of contacts removed from non-household layers for the simulations with contact reduction. The y-axis shows the total number of infections, averaged over 100 runs.}
    \label{fig:asym_cases}
\end{figure*}

\begin{figure*}
    \centering
    \begin{subfigure}{0.5\textwidth}
        \centering
        \includegraphics[width=\textwidth]{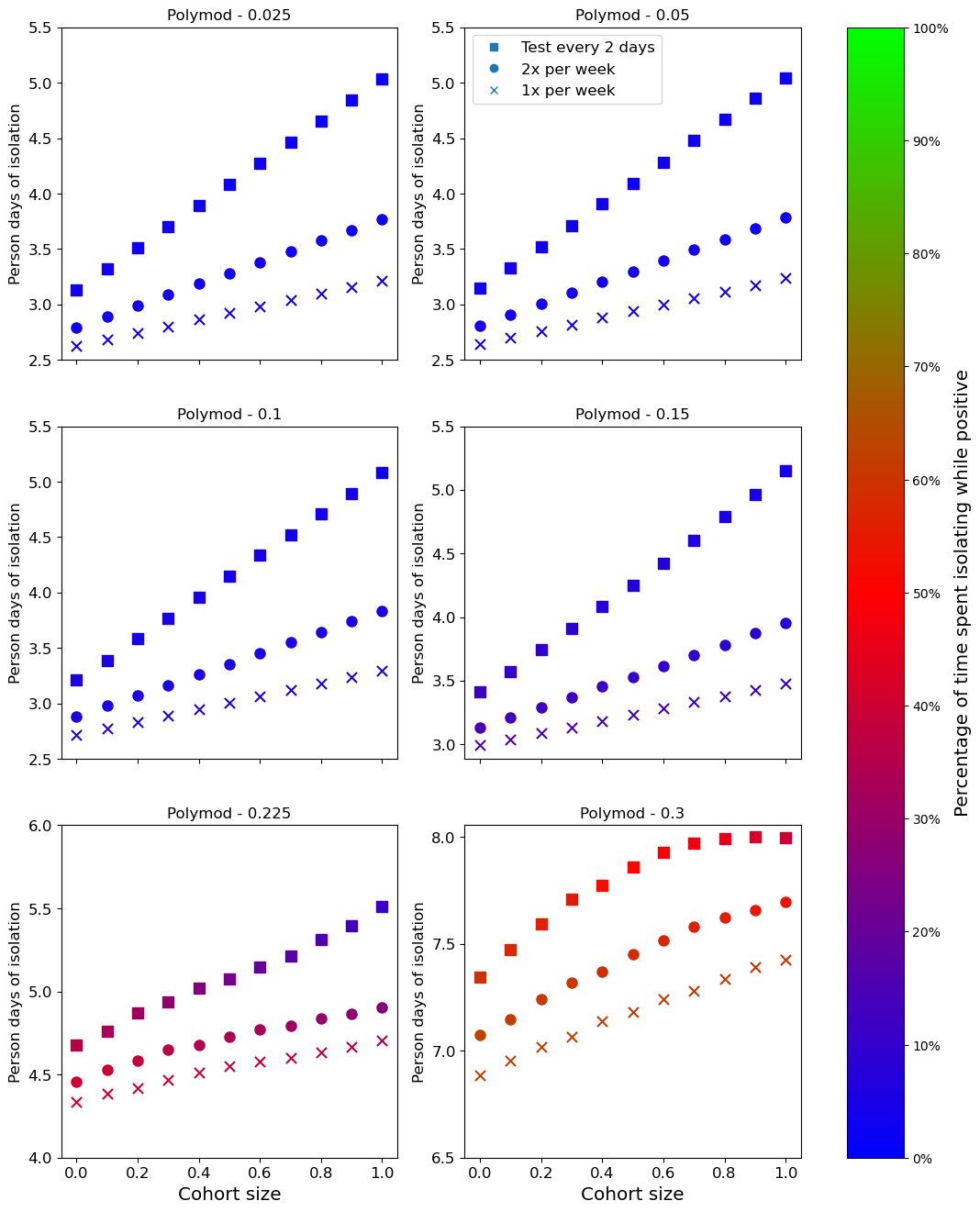}
        \caption{Contact patterns obtained from POLYMOD}
        \label{fig:asym_isolation_polymod}
    \end{subfigure}%
    \begin{subfigure}{.5\textwidth}
        \centering
        \includegraphics[width=\textwidth]{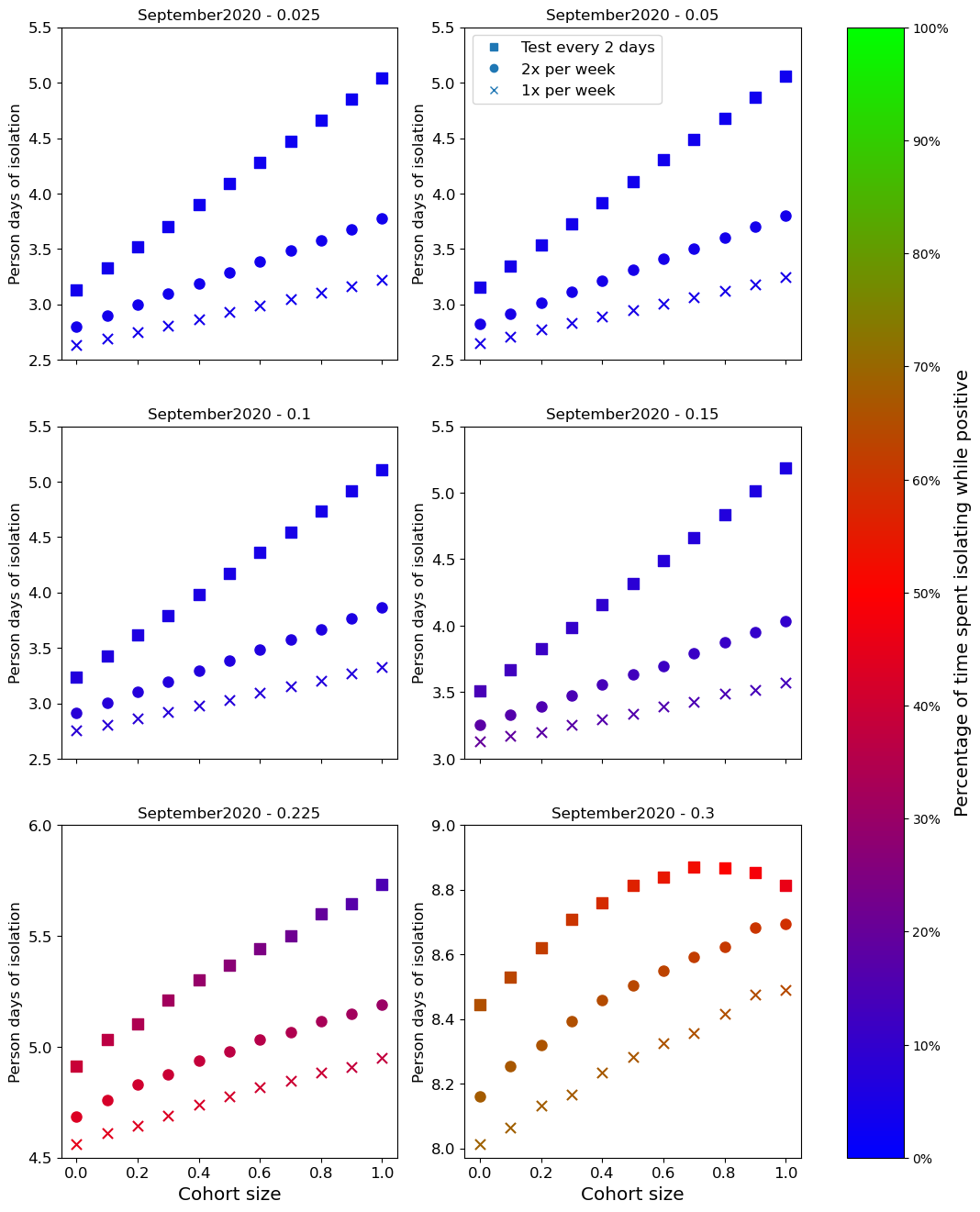}
        \caption{Contact patterns obtained from CoMix round 22 to 27 (late August 2020 to September 2020)}
        \label{fig:asym_isolation_comix}
    \end{subfigure}
    \caption{Impact on person-days of isolation of increasing adherence to the different asymptomatic testing strategies, for different growth rates and contact patterns. The x-axis represents the size of the regular tester cohort for the simulations with asymptomatic testing or the percentage of contacts removed from non-household layers for the simulations with contact reduction. The y-axis shows the mean number of person days of isolation per population members over the 180 days of simulation, averaged over 100 runs. These include days in isolation waiting for test results (amongst true positives and true negatives); isolation days amongst those testing positive who were true positive; and isolation days among those testing positive who were false positives. The colour of each marker indicates the percentage of days spent in isolation while infected.}
    \label{fig:asym_isolation}
\end{figure*}

We show the comparison of these two approaches through the number of total
infections after the 180 days of simulation, in Figure~\ref{fig:asym_cases},
and the number of person days of isolation, in Figure~\ref{fig:asym_isolation}. We also show the relationship between the size of the peak in number of daily infections and the number of total infections, in the Appendix (Figure~\ref{fig:asym_peak_cases}).

For a twice weekly testing policy, which was the government advice for regular
asymptomatic testing during 2021, we find that at 100\% adherence, the decrease
in total number of infections compared to 0\% adherence to this policy is
within the range [6.6\%, 44.6\%] for POLYMOD contact patterns and [7.0\%,
44.9\%] for CoMix contact patterns, across the different growth rates, as shown in
Figure~\ref{fig:asym_cases}. The growth rates at which these boundary values
occur are 0.025 and 0.15 for both POLYMOD and CoMix. This reduction in the
number of total infections is similar to the impact of reducing the number of
contacts by 25 to 30\% across all growth rates, for both sets of contact
patterns. At 50\% adherence, these intervals are reduced to [3.3\%, 27.9\%] for
POLYMOD and [4.1\%, 26.0\%] for CoMix.

For a once weekly testing policy, where there is 100\% adherence, the decrease in total
number of infections ranges from 3.9\% at 0.025 growth rate to 32.2\% at 0.15
with POLYMOD contact patterns, and from 3.9\% at 0.025 growth rate to 32.9\% at
0.15 with CoMix contact patterns. The decline in the total number of infections
is equivalent to reducing the number of contacts by 15 to 20\%, across all
growth rates and for both sets of contact patterns. At 50\%
adherence, this decrease is reduced to 1.6\% at 0.025 growth rate and
18.4\% at 0.15 with POLYMOD contact patterns, and to 2.7\% at 0.05 growth rate
and 19.9\% at 0.15 with CoMix contact patterns.

For the most intensive asymptomatic testing policy, testing every 2 days, given
100\% adherence, the decrease in total number of infections is within the range
[8.6\%, 71.0\%] for POLYMOD contact patterns and [7.7\%, 65.2\%] for CoMix
contact patterns. The growth rates at which these boundary values occur are
0.025 and 0.15 for both contact patterns. This reduction is equivalent to the
impact of reducing the number of contacts by 40 to 45\% for all growth rates,
in both sets of contact patterns. At 50\% adherence, these intervals are
reduced to [4.6\%, 40.9\%] for POLYMOD and [4.0\%, 37.9\%] for CoMix.

We find that the impact of testing on the maximum number of infections per day
is approximately the same as the impact on number of total infections, when
compared to reducing the number of contacts, shown in Figure~\ref{fig:asym_peak_cases}.
Results show that testing every 2 days yields a peak height equivalent to
reducing the number of contacts by 40 to 45\%, testing twice per week is
equivalent to reducing contacts by 20 to 25\% and testing once a week is
equivalent to reducing contacts by 10 to 15\%. These numbers are consistent
across multiple growth rates and contact patterns.

Analysis of person-days of isolation reveals similar behaviour when the growth
rate is less than or equal to 0.1, for both contact patterns,
Figure~\ref{fig:asym_isolation}. In these low prevalence cases, increasing the
adherence to the testing policy leads to a increase of 22.5\% (0.6 days), 35\%
(1 day) and 60\% (1.9 days) in number of person-days of isolation for the
weekly, twice weekly and testing every two days policies, respectively. The
time spent in isolation while positive accounts for less than 10\% of the time
in isolation, for 0.025, 0.05 and 0.1 growth rates.

For higher growth rates, increasing testing leads to a smaller increase in the
number of person-days of isolation, with the increase becoming smaller the
higher the growth rate is. For the testing every two days policy, increasing
adherence to 100\% leads to an increase in number of person days of isolation
of 51\% (1.7 days), 17.7\% (0.9 days) and 8.9\% (0.6 days) for 0.15, 0.225 and
0.3 growth rates, respectively, when using POLYMOD contact patterns, and 47\%
(1.6 days), 16.6\% (0.8 days) and 4.3\% (0.4 days), when using CoMix contact
patterns. For the twice weekly testing policy, increasing adherence to 100\%
leads to an increase in number of person days of isolation of 26\% (0.8 days),
10.1\% (0.4 days) and 8.8\% (0.6 days) for the same growth rates, when using
POLYMOD contact patterns, and 23.9\% (0.8 days), 10.8\% (0.5 days) and 6.5\%
(0.5 days), when using CoMix contact patterns.

For all growth rates, this increase in isolation time is driven by an increase
in people isolating from false positive LFD tests. This is supported by
inspecting the proportion of time spent isolating while positive with COVID,
which decreases as adherence to testing increases, represented by the colour
gradient in Figure~\ref{fig:asym_isolation}. This effect is more pronounced at
high growth rates and higher rates of testing. For instance, at a growth rate
of 0.3 with POLYMOD contact patterns, for the testing every two days policy,
the time spent isolating while positive varies from 60\% at 0\% adherence to
42\% at 100\%, but for the twice weekly testing policy, these values change to
63\% and 54\%, respectively.

\subsection{Symptomatic testing}
\label{sec:sympt}

\begin{figure*}[h]
    \centering
    \begin{subfigure}{0.85\textwidth}
        \centering
        \includegraphics[width=\textwidth]{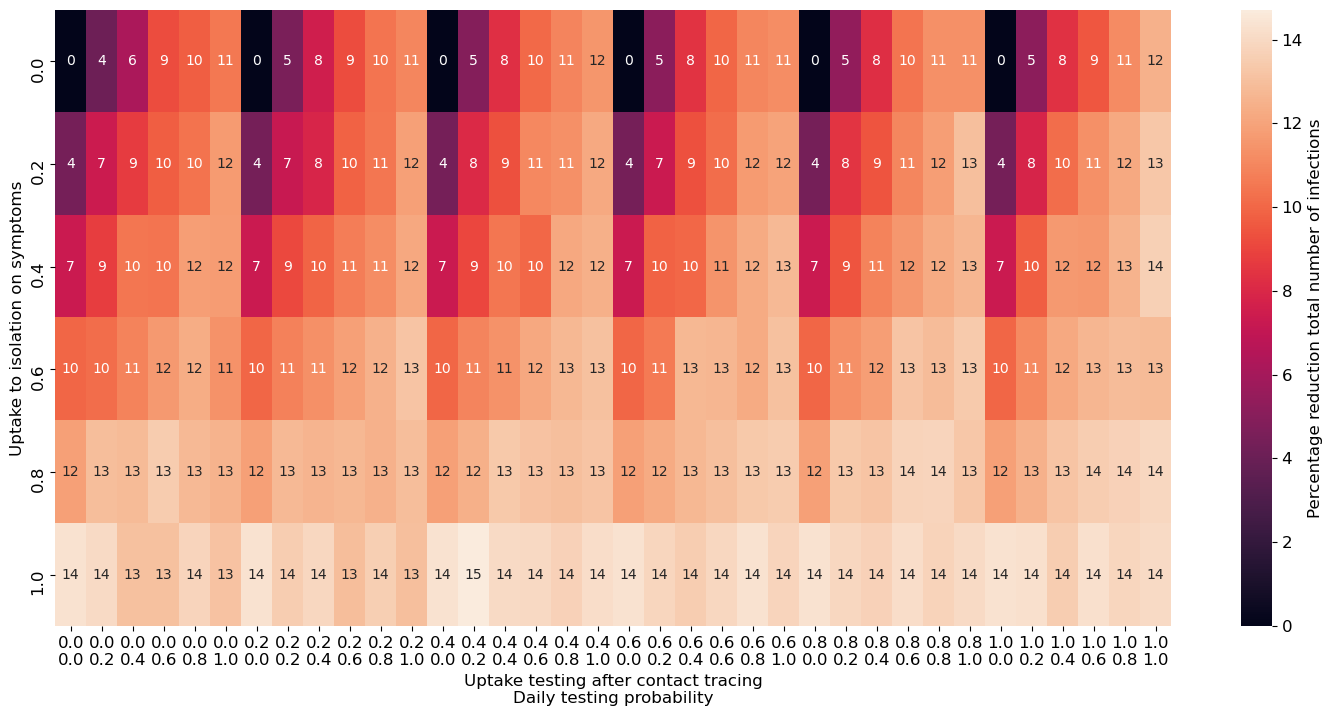}
        \caption{Growth rate = 0.025}
        \label{fig:symp_cases_0.025}
    \end{subfigure}

    \begin{subfigure}{.85\textwidth}
        \centering
        \includegraphics[width=\textwidth]{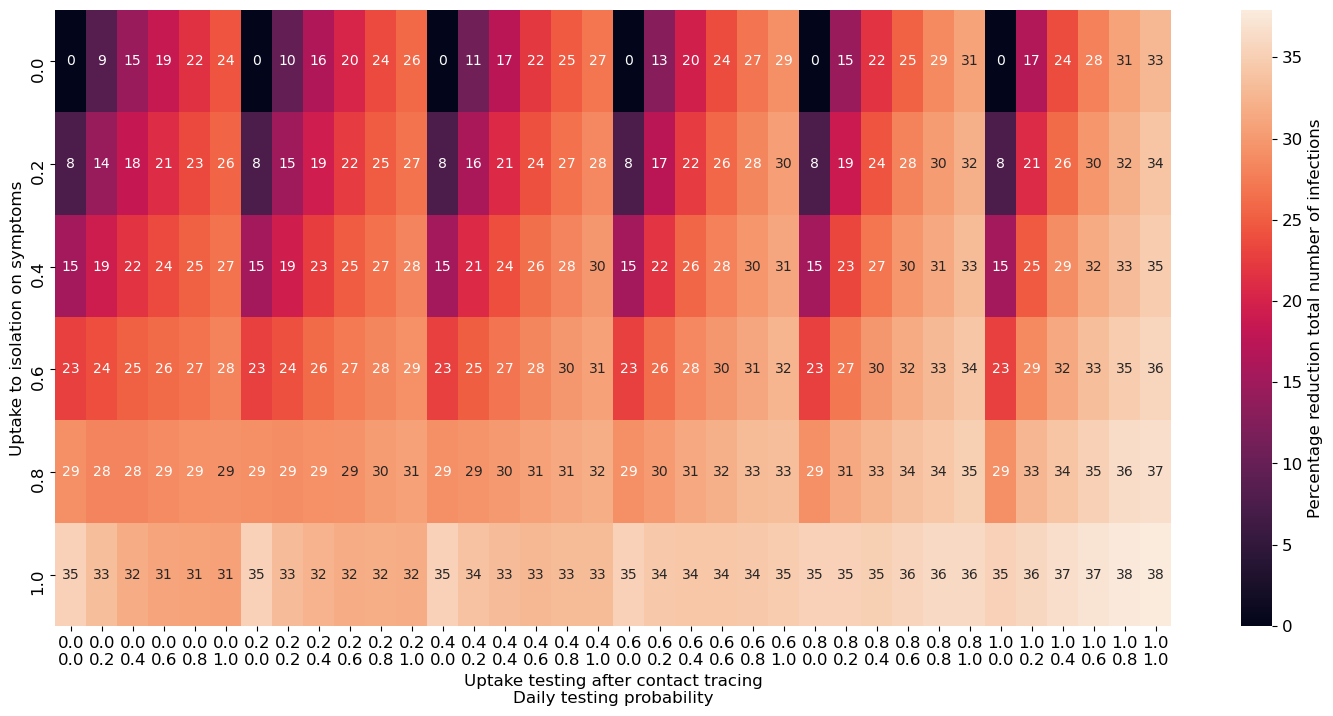}
        \caption{Growth rate = 0.3}
        \label{fig:symp_cases_0.3}
    \end{subfigure}
    \caption{Impact of increasing the uptake on isolation on symptoms, daily symptomatic testing and PCR testing after being traced as a contact of a positive case. Numbers shown are the percentage decrease in the total number of infections, compared to 0\% uptake on these interventions, for 0.025 and 0.3 growth rates in POLYMOD contact pattern. The top line on the x axis represents the uptake to testing after contact tracing, the bottom line represents the daily probability to testing when symptomatic. The y axis represents uptake to isolation on COVID or ILI symptoms.}
    \label{fig:sym_cases}
\end{figure*}

We also explore how sensitive our model is to changes in the parameters that
control adherence to isolation on symptoms, daily probability of testing if
symptomatic and the proportion of contacts that take up a PCR. We vary these
parameters in increments of 20\% from 0\% to 100\%. An uptake of 100\% in these
interventions means that all traced contacts take a PCR test when they are
traced and everyone that develops COVID or ILI symptoms isolates and performs a
PCR test the day after isolating, continuing to isolate for a total of 10 days
if the test is positive and being released from isolation if the test is
negative. These experiments are conducted solely on POLYMOD contact patterns,
for the multiple growth rates as above, due to the heavy computational load.
The impact of changing these uptake parameters in the total number of
infections is shown in Figure~\ref{fig:sym_cases}.

Results show that the most effective of three interventions at reducing
transmission is isolation on symptom onset, across all growth rates. This is a
very intuitive result and follows our expectation, as isolation naturally
prevents most transmission. Perfect adherence to isolation on symptoms with no
contact tracing and no testing accounts for a reduction in number of cases
between 14.3\% (at a 0.025 growth rate) and 66.9\% (at a growth rate of 0.15).

On the other hand, contact tracing and asking contacts to take a PCR test,
isolating only if positive, is the least effective of these interventions. With
a 100\% testing probability when symptomatic and no isolation on symptoms, the
combination of parameters that should favour contact tracing the most as
contacts can only be traced when there is testing, our results show that
contact tracing reduces the number of cases between 1.9\% (at a growth rate of
0.025) and 14.5\% (at a growth rate of 0.225).

The impact of symptomatic and contact testing is harder to measure, as it
interacts with the effectiveness of the other two interventions. If we assume
testing but no isolation on symptoms (only after a positive test) and no
contact tracing, testing can reduce the number of cases from 10.6\% (at a
growth rate of 0.025) to 51.6\% (at a growth rate of 0.15). However, if we
assume perfect adherence to isolation on symptoms, the impact of testing
depends on the adherence to testing after being contact traced. For example, at
a growth rate of 0.15, with no contact tracing, increasing the daily
probability of testing from 0 to 100\% leads to more cases, from an average of
5340 to 6235, corresponding to a reduction of 66.9\% to 61.3\% when compared to
no interventions. But if we add contact tracing with perfect adherence,
increasing the testing probability leads to less cases, from an average of 5340
to 5036, corresponding to a reduction of 66.9\% to 68.7\% when compared to no
interventions.

The value of 40\% adherence to the 3 interventions we chose for the
asymptomatic experiments, correspond to a reduction in number of cases by
10.5\%, 15.6\%, 39.3\%, 50.2\%, 29.6\% and 23.9\%, for each growth rate,
respectively. These figures also indicate that testing is most effective when
the epidemic is at a medium level of growth; if the epidemic is at a very high
rate of growth testing is not enough to curb transmission.

\subsection{Comparing PCR and LFD tests}
\label{sec:lfd_vs_pcr}

\begin{figure*}
    \centering
    \begin{subfigure}{0.5\textwidth}
        \centering
        \includegraphics[width=\textwidth]{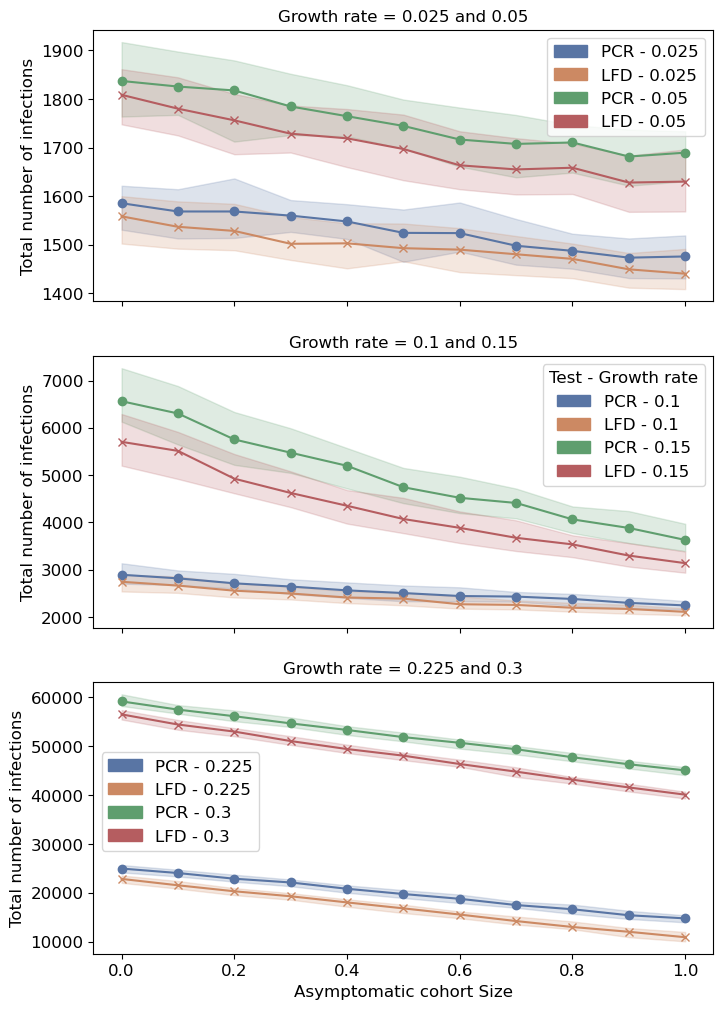}
        \caption{Asymptomatic testing twice per week}
        \label{fig:asym_lfd_vs_pcr_twiceperweek}
    \end{subfigure}%
    \begin{subfigure}{.5\textwidth}
        \centering
        \includegraphics[width=\textwidth]{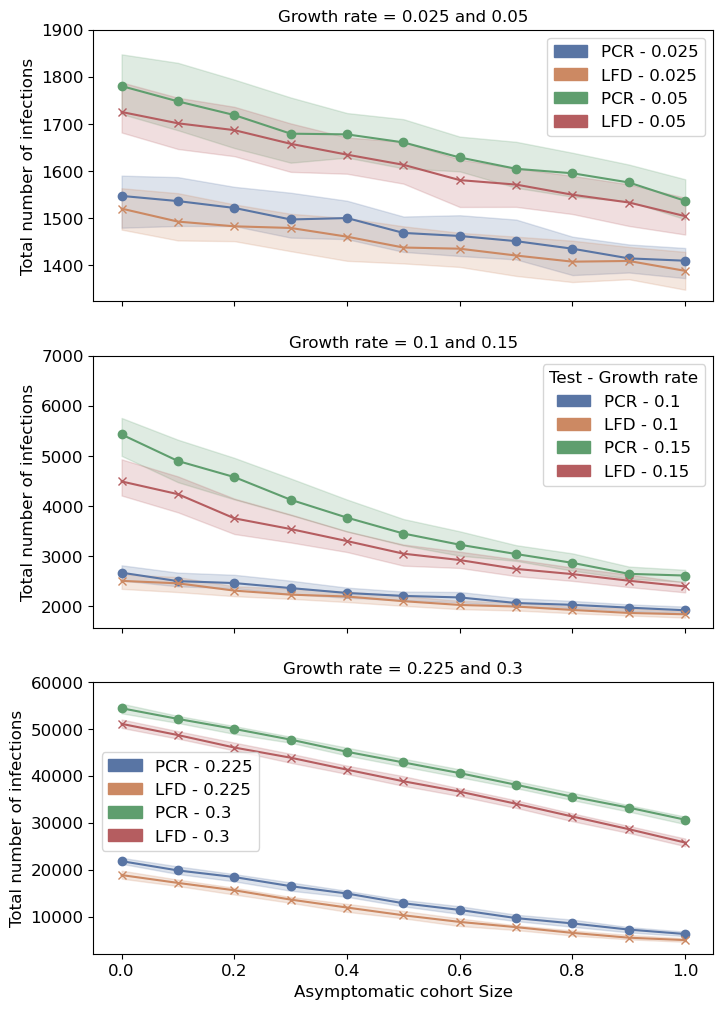}
        \caption{Asymptomatic testing every two days}
        \label{fig:asym_lfd_vs_pcr_every2days}
    \end{subfigure}
    \caption{Comparison of LFD against PCR tests for symptomatic or contact traced individuals, for different values of asymptomatic testing uptake and different growth rates, with contact patterns obtained from POLYMOD. The y-axis shows the total number of infections. The x-axis represents the size of the regular asymptomatic tester cohort. Markers shown represent the median of 100 simulations and the shaded area represents the interquartile range. Figure~\ref{fig:asym_lfd_vs_pcr_twiceperweek} corresponds to the twice weekly testing policy and Figure~\ref{fig:asym_lfd_vs_pcr_every2days} to the testing every two days policy.}
    \label{fig:asym_lfd_vs_pcr}
\end{figure*}

We now explore how substituting PCR tests - taken by symptomatic individuals
and by contacts of identified cases - with LFD tests impacts the total
number of infections, visualized in Figures~\ref{fig:asym_lfd_vs_pcr}
and~\ref{fig:sym_lfd_vs_pcr}. These results all use POLYMOD contact patterns.

We find that using an LFD test in
lieu of a PCR test among symptomatic individuals and their traced contacts yields a
smaller total number of infections, with particular significance at high growth
rates of the epidemic. For a testing every two days policy, the difference
ranges from 1.6\% (0.025 growth rate) to 20\% (0.225 growth rate) when the
whole population is testing frequently and from 1.9\% (0.025 growth rate) to
19\% (0.225 growth rate) when half the population is in the regular testing group.
For a twice weekly testing policy, the difference ranges from 1.9\% (0.025
growth rate) to 25\% (0.225 growth rate), when the whole population is testing
regularly, and from 1.7\% to 16\%, at the same growth rates, when half the
population is testing regularly.

In the experiments with no asymptomatic testing, we find that at small growth
rates, there is no significant difference between using a PCR versus single LFD
test for symptomatic individuals and contacts. We find that at medium growth
rates (0.1 and 0.15, the latter visualized in Figure~\ref{fig:sym_lfd_vs_pcr}),
across different values for adherence to isolation on symptom onset and daily
probability of testing, using LFD tests leads to a smaller number of infections
if the adherence to testing after contact tracing is less than 40\%, after
which PCR leads to on average to a lower number of infections. Finally, results
show that at high growth rates (0.225 and 0.3), LFD tests consistently lead to
a smaller number of infections, in particular when the probability of testing
if symptomatic is 100\%, where the average (across the range of uptakes to testing
after contact tracing) difference between PCR and LFD tests is 7.6\%, 6.6\% and
5.0\% with no isolation on symptoms, 40\% adherence and 100\% adherence,
respectively.

Another metric of interest in this context is the number of tests required to
implement this increase in adherence to testing, symptomatic or asymptomatic.
In the asymptomatic case, we find no difference in the total number of tests
required between using PCR or LFD tests for symptomatic individuals and
contacts, which is expected because the majority of testing is done in
asymptomatic individuals. Implementing the asymptomatic strategies if the whole
population took up asymptomatic testing requires 30.1, 49.2 and 91.9 tests per
person for the weekly, twice weekly and testing every 2 days policies,
respectively. These correspond to increases of 180\% (19.4 extra tests), 201\%
(32.9 extra tests) and 235\% (64.4 extra tests) over no one in the population
taking up the regular asymptomatic tests.

For the symptomatic experiments, we show the average number of tests per person
taken over the 180 days of simulation in Figure~\ref{fig:lfd_vs_pcr_tests}.

Results consistently show across different growth rates that more LFD tests are
used on average per person. An exception is when the growth rate is equal to
0.3: due to the significantly greater number of infections when using PCR
tests, more tests of this type are required as more people become symptomatic and
more people are contact traced. This increased requirement for PCR tests is
more noticeable at high rates of adherence to testing after contact tracing: at
100\% adherence to testing after contact tracing there are on average 66.3\%
more PCR tests consumed per person, whereas at 0\% adherence there are on
average 34.1\% more LFD tests used per person.

Swapping PCR tests for LFTs also affects the amount of time each person spends
in isolation, as shown in
Figure~\ref{fig:asym_pcr_vs_lfd_isolation} for experiments with asymptomatic
testing and Figure~\ref{fig:sym_pcr_vs_lfd_isolation} for experiments with symptomatic testing
only.

In the asymptomatic case, we find that the difference between both tests is
similar for 0.025, 0.05, 0.1 and 0.15 growth rates; people isolate longer on
average when LFD tests are used, due to the possibility of a false negative. In
these growth rates, with 100\% adherence to the testing policy, people spend on
average 17.9\% (0.6 days), 14.0\% (0.5 days) and 7.8\% (0.4 days) longer in
isolation if LFD tests are used for the weekly, twice weekly and testing every
2 days policies, respectively. This difference is reduced when the growth rate
equals 0.225, to 2.9\% (0.1 days), 2.4\% (0.1 days) and 4.3\% (0.2 days) and
reversed when the growth rate is 0.3, leading to people isolating longer when
PCR tests are used by 6.6\% (0.5 days), 5.9\% (0.4 days) and 5.9\% (0.4 days).
For the 0.3 growth rate, the longer isolation times when using PCR tests are
caused by a greater number of infections, with the difference being large
enough to be reflected in the number of person-days of isolation.

At higher levels of adherence to regular asymptomatic testing, the increase in
the number of person-days of isolation caused by using an LFD test instead of a PCR is
smaller than at lower levels of adherence. For example, for the twice weekly
testing policy, where there are 0.025, 0.05, 0.1 and 0.15 growth rates, the average increase
in isolation days when using PCR tests rather than LFTs is 21.7\% (0.6 days) at
0\% adherence and 17.4\% (0.6 days) at 50\% adherence. This is a consequence of
the greater number of false positives resulting from intensive asymptomatic testing at
high adherence, with the majority of isolation time being a consequence of
asymptomatic testing. Switching to LFD testing in symptomatic (ILI or COVID)
and contact traced individuals makes less difference to this metric under these circumstances.

Results also show that when using LFTs to test symptomatic or contact traced
individuals, the proportion of time spent isolating while infected with COVID
is smaller across all growth rates and testing strategies. Using the same
example as before, at a growth rate of 0.3 with POLYMOD contact patterns, for
the testing every two days policy, the time spent isolating while positive
varies from 49\% at 0\% adherence to 31\% at 100\%, and for the twice weekly
testing policy, these values change to 51\% and 44\%, respectively. This is
likely to be caused by a combination of better infection control (fewer
infections lead to less time isolating while positive) and more false
positives (which lead to more time isolating while negative for COVID).

A similar pattern is observed in the symptomatic experiments, where the
number of person-days in isolation when testing with PCR or
LFD is similar for 0.025, 0.05, 0.1 and 0.15 growth rates. We observe that
contact tracing only affects the number of person-days in isolation when the
growth rate is 0.3, where increasing adherence to testing after contact
tracing using PCR tests leads to an average of 0.8 extra person-days in
isolation, against 0.2 days when using LFD tests. This contrasts with
the remaining growth rates, where increasing adherence leads to less than 0.1
extra person-days in isolation.

Results also show that when increasing adherence to isolation on symptom-onset,  using LFD tests leads to more
person-days of isolation than using PCR tests. Under the 0.025, 0.05, 0.1
and 0.15 growth rates, people spend on average 0.1 extra days in isolation
when using LFTs than PCRs, when adherence to isolation on symptom-onset is 0\%,
growing to 1.4 days when adherence is 100\%. Under the 0.225 and 0.3 growth
rates, at low levels of adherence to isolation on symptoms, people spend more
time in isolation when PCR tests are used, 0.5 days at 0.225 growth rate and
1.5 days at 0.3 growth rate. As adherence to isolation on symptom-onset increases, this value shifts so
that using LFD tests leads to people spending more time in isolation, 1.0 extra
days at a 0.225 growth rate when adherence is 100\% and 0.4 days at a 0.3 growth
rate.

Over all, the greater the adherence to isolation on symptom onset, the less
time spent isolating while negative to COVID. This highlights how increasing
adherence to isolation on symptom onset achieves a high reduction in the number of
infections: by preemptively isolating individuals with any symptoms that could
be related to COVID, many susceptible people are being removed from the contact
networks and are therefore less likely to be infected.

Similarly to the asymptomatic experiments, when using LFD tests individuals
spend more time isolating while they are negative to COVID. Excluding when
there is no isolation on symptom onset, when using PCR tests individuals spent
on average 5\% of their time in isolation positive to COVID, with a growth of
0.025, up to 56\%, with a growth rate of 0.3. When using LFD tests instead,
these values are reduced to 3 and 47\%, respectively.

\section{Discussion}

We looked at three different policies for asymptomatic testing, one based on UK government advice in Autumn 2021 that recommended LFD testing twice per week and two alternatives that doubled and halved the recommended frequency of testing. These asymptomatic testing policies are compared against a scenario of partial closure of schools and work places and reductions in community contacts, which could be implemented through a dissuasion of activities that require indoor mixing or placing limits on groups sizes in restaurants/pubs. These scenarios emulate lockdowns at different levels and are naturally very effective at infection control. However, they carry a high cost in the form of economic and mental health impact on citizens that we do not account for in the simulations, and it may be unrealistic to maintain these measures of infection control for as long as 6 months (the length of our simulations).

Even with an experimental setup that favours a reduction of contacts, we find that with the twice weekly testing schedule, increasing the number of people participating in regular asymptomatic testing can have as much impact as reducing the number of contacts by 30\%. When halving/doubling the testing frequency this impact is reduced/increased to up to 20\% or 45\%, respectively.

Asymptomatic testing strategies carry a cost that can be measured by the number of tests each person takes and by how long each person spends in isolation. Perfect adherence to the twice weekly asymptomatic testing advice requires that on average each person in the whole population take approximately 50 tests over a 180 day period, the majority of which are LFD, and spend approximately 3.5 days isolating while not infected with COVID-19. These days spent in isolation are, for the most part, a consequence of preemptive isolation while waiting for a test result, due to developing symptoms (COVID or ILI). The increase in testing, which identifies more infections but also carries a risk of generating more false positives, is only responsible for an increase of less than 1 person day in isolation.

We also explored the impact of different levels of adherence to infection control policies in a scenario where there is no asymptomatic testing in the population. In this situation, we find that the most effective policy for decreasing the number of infections is preemptive isolation on symptom onset, while waiting for a test result. This highlights the need for widespread education about the most common symptoms of COVID, particularly if new variants emerge that show different profiles of symptoms, and the government's support to facilitate this isolation period. It is also necessary to support high levels of testing to prevent long periods of isolation for people with ILI symptoms that overlap with COVID.

However, where there is little or no appetite to support preemptive isolation periods, our experimental setup shows that high rates of symptomatic testing and adherence to testing after contact tracing can achieve a similar reduction in the number of infections in the population. In fact, our assumptions about the percentage of contacts found through contact tracing are conservative according to estimates from UK NHS Test and Trace \citep{nhsTT}, so the effectiveness of the combination of symptomatic testing and contact tracing could be higher than the results of our simulations, especially considering that achieving higher adherence to these policies is easier than high adherence to preemptive isolation.

Finally, we studied the effect of replacing PCR tests with LFTs in symptomatic and contact traced individuals and show that in the majority of scenarios, this substitution leads to a smaller number of total infections. This difference is caused by the testing delay associated with PCR tests, as the need to develop the test results in a laboratory setting offsets the gain obtained by being more accurate at identifying positive individuals. The greater sensitivity of PCR tests is even less important when we factor in the ability to perform multiple LFD tests before learning the result of a PCR test, especially because the majority of tests done in this context are by symptomatic individuals (rather than contact traced ones) so it becomes imperative to diagnose and isolate these individuals when they are at the peak of their infectivity.

The reduction in number of infections achieved by replacing PCR tests with LFTs is achieved by a trade-off that can be quantified through an increase in both the number of tests done and the average number of person days in isolation; the latter is caused by false positive tests, leading to people spending a greater portion of their isolation periods not infected with COVID. The increase in the number of tests when using LFD rather than PCR is only noticeable in the experiments with no asymptomatic testing and is a consequence of the lower test sensitivity; individuals are more likely to test negative with an LFD test than with a PCR but, because they are symptomatic, they are likely to take more tests during their symptomatic period. This increase in the number of tests is not enough to offset the cost disparity between PCR and LFD tests, as each LFD test is much cheaper than a PCR test and requires less infrastructure.

Our analysis has a number of strengths. Our results are similar for different contact patterns, implying that our conclusions do not require  the assumption that average number of contacts stays as low as during the pandemic. We examine a range of growth rates to cover different scenarios of epidemic growth, allowing us to study the cases where no interventions are needed (growth rates of 0.025 and 0.05), when testing is most effective (growth rates of 0.1, 0.15 and 0.225) and when testing alone is not sufficient (growth rate of 0.3).

Our work has some limitations. Firstly, Covasim is a stochastic model and the inherent stochasticity gives rise to uncertainty in addition to uncertainty from the model parameters. For the purposes of this analysis we ran 100 seeds for each set of parameters, to give sufficient robustness in generating central, low and high outcomes across the scenarios. Future work will look at how to better quantify uncertainty across agent based models such as Covasim.

Secondly, in this study and when using Covasim in general, while most of the parameters are derived from the literature, there are gaps where we had to make assumptions. For example, for the asymptomatic experiments, we assumed 40\% background symptomatic testing, adherence to isolation on symptoms and adherence to testing after contact tracing. Although this assumption is based on the results of the symptomatic experiments, it is unclear the extent to which the amount of symptomatic testing affects the efficacy of asymptomatic testing. Another assumption in the asymptomatic experiments is how much asymptomatic testing people outside the regular testing cohort should perform. A sensitivity analysis on the chosen parameters (90/10\% split for intermittent and anti testers) would be useful to understand the impact of this on our results.

Similarly, while we have made an effort to quantify vaccine efficacy against onward transmission of different variants and waning protection from both vaccination and infection, current data is not definitive about these. Specifically, the version of Covasim used in this study assumed a single variant (albeit with different step-like change in transmissibility) and a single antibody waning function for all individuals and all types of immunity, with individual- and immune-level variation in the level of Neutralising Antibodies (NAbs). Incorporating different --possibly competing-- variants and different NAbs trajectories was beyond the scope of this work. Finally, we calibrated the model to aggregated national data as this was sufficient for this study. Future work will look at extending the granularization of the model by both age and region.

\section{Conclusion}

We explored the impact of different testing strategies using the Covasim agent-based model. Our results highlight that asymptomatic testing with LFTs is particularly effective when the growth rate corresponds to a weekly doubling in number of cases. We also found that swapping PCR tests for LFD may help in epidemic control due to the delay in returning test results to individuals, which seems to offset the loss in test sensitivity in LFD.

Overall our findings highlight that regular asymptomatic testing with LFTs can be a viable alternative to national lockdowns. Delivering a strategy of large scale testing with LFTs combined with immediate self-quarantining and tracing of contacts will be crucial in dampening transmission, reducing hospitalisations and deaths in any future COVID-19 outbreaks as countries move towards a “living with COVID-19” strategy.

\section*{Acknowledgements}

This work was funded by United Kingdom Research and Innovation Medical Research Council grant number MR/V028618/1.

MEPS is funded by Engineering and Physical Sciences Research Council Manchester Centre for Doctoral Training in Computer Science (grant number EP/I028099/1). M.F. is supported by the Alan Turing Institute under the EPSRC grant no. EP/N510129/1. JPG's work was supported by funding from the UK Health Security Agency and the UK Department of Health and Social Care (DHSC). TH and LP are supported by the UKRI through the JUNIPER modelling consortium (grant no. MR/V038613/1). TH is also supported by the
Engineering and Physical Sciences COVID-19 scheme (grant number EP/V027468/1), the Royal Society (grant number INF/R2/180067) and the Alan Turing Institute for Data Science and Artificial Intelligence.

\printcredits


\bibliographystyle{model1-num-names}

\bibliography{paper}

\clearpage
\appendix

\section{Covasim Parameters}
\label{app:parameters}

\begin{table*}[h!]
    \centering
  \renewcommand{\arraystretch}{1.5}
    \begin{tabular}{|>{\centering\arraybackslash}m{80pt}|>{\centering\arraybackslash}m{280pt}|>{\centering\arraybackslash}m{90pt}|}
    \hline \textbf{Parameter name} & \textbf{Parameter description} & \textbf{Value} \\
    \hline Population size & Number of agents in the simulation & 100,000 \\
    \hline Initial infections & The number of agents initially infected & 100 \\
    \hline Number of days & Number of time steps, i.e., length of the simulation & 180 \\
    \hline Number of imports & Number of daily new infections from a source outside the population of study & Pois(5) \\

    \hline \multicolumn{3}{c}{\emph{Testing}}\\
    \hline Rapid Antigen Test (LFD) specificity & Probability of a negative test given that the person is not infected & 99.7\% \citep{wolf2021lateral} \\
    \hline Bad swab rate & Proportion of PCR and LFD tests returning negative regardless of infection status & 10\% \\
    \hline Test sensitivity & Probability of a positive test given that the person is infected & Test sensitivity curves based on~\cite{hellewellestimating} \\
    \hline PCR test delay & Number of days for the test results to be known & Pois(1.2) \\

    \hline \multicolumn{3}{c}{\emph{Isolation}}\\
    \hline ILI prevalence & Percentage of the population infected with ILI symptoms per day. These symptoms persist for $\text{Lognormal}(6.0,2.0)$ days & 1\% \\
    \hline Isolation period & Number of days someone isolates after testing positive & 10 \\
    \hline Isolation factor & Percentage reduction of transmission rate parameter per layer while in isolation & Household: 0\%, School and Work: 90\%, Community: 90\% \\

    \hline \multicolumn{3}{c}{\emph{Contact tracing}}\\
    \hline Contact tracing probability & Proportion of contacts traced per layer after a positive test & Household: 100\%, School and Work: 45\%, Community: 15\% \\
    \hline Contact tracing time & Number of days for contacts to be traced & Household: 0, School and Work: 1, Community: 2 \\
    \hline
    \end{tabular}
    \caption{Table of parameters modified from Covasim defaults.}
    \label{tab:def_params}
\end{table*}

\clearpage

\section{Summary of experimental setup}

\begin{table*}[hb]
    \centering
  \renewcommand{\arraystretch}{1.7}
    \begin{tabular}{|>{\centering\arraybackslash}m{80pt}|>{\centering\arraybackslash}m{280pt}|>{\centering\arraybackslash}m{90pt}|}
    \hline \textbf{Parameter name} & \textbf{Parameter description} & \textbf{Value} \\
    \hline \multicolumn{3}{c}{\emph{Asymptomatic testing}}\\
    \hline Testing frequency & How often people in the regular tester cohort test & Every 2 days; twice per week; once per week \\
    \hline Cohort size & What percentage of the population is testing frequently on LFD tests & Varies from 0 to 100\% in jumps of 10\% \\
    \hline Anti testing group size & What percentage of the population not in the regular testing group that refuses to test without any symptoms & 10\% \\
    \hline Intermittent testing group size & What percentage of the population not in the regular testing group that tests at a quarter of the rate of the regular testers & 90\% \\
    \hline Isolation on symptom onset & Adherence to isolation when developing COVID-19 or ILI symptoms & 40\% \\
    \hline Daily testing probability if symptomatic & Probability of taking up a test per day if symptomatic with COVID-19 or ILI symptoms & 40\%; test is PCR in Section~\ref{sec:asympt} or LFD in Section~\ref{sec:lfd_vs_pcr} \\
    \hline Test uptake after contact tracing & Adherence to taking up a test when notified by tracing of a positive contact & 40\%; test is PCR in Section~\ref{sec:asympt} or LFD in Section~\ref{sec:lfd_vs_pcr} \\

    \hline \multicolumn{3}{c}{\emph{Reduction of number of contacts}}\\
    \hline Contact reduction & Proportion of non-household contacts removed from the contact networks & Varies from 0 to 100\% in jumps of 5\% \\
    \hline Isolation on symptom onset & Adherence to isolation when developing COVID-19 or ILI symptoms & 40\% \\
    \hline Daily testing probability if symptomatic & Probability of taking up a PCR test per day if symptomatic with COVID-19 or ILI symptoms & 40\% \\
    \hline Test uptake after contact tracing & Adherence to taking up a PCR test when notified by tracing of a positive contact & 40\% \\

    \hline \multicolumn{3}{c}{\emph{Symptomatic testing}}\\
    \hline Isolation on symptom onset & Adherence to isolation when developing COVID-19 or ILI symptoms & Varies from 0 to 100\% in jumps of 20\% \\
    \hline Daily testing probability if symptomatic & Probability of taking up a test per day if symptomatic with COVID-19 or ILI symptoms & Varies from 0 to 100\% in jumps of 20\%; test is PCR in Section~\ref{sec:sympt} or LFD in Section~\ref{sec:lfd_vs_pcr} \\
    \hline Test uptake after contact tracing & Adherence to taking up a test when notified by tracing of a positive contact & Varies from 0 to 100\% in jumps of 20\%; test is PCR in Section~\ref{sec:sympt} or LFD in Section~\ref{sec:lfd_vs_pcr} \\
    \hline
    \end{tabular}
    \caption{Table of parameters we vary in each experiment.}
    \label{tab:exp_params}
\end{table*}

\clearpage

\section{Additional figures}

\begin{figure*}[!h]
    \centering
    \begin{subfigure}{0.5\textwidth}
        \centering
        \includegraphics[width=\textwidth]{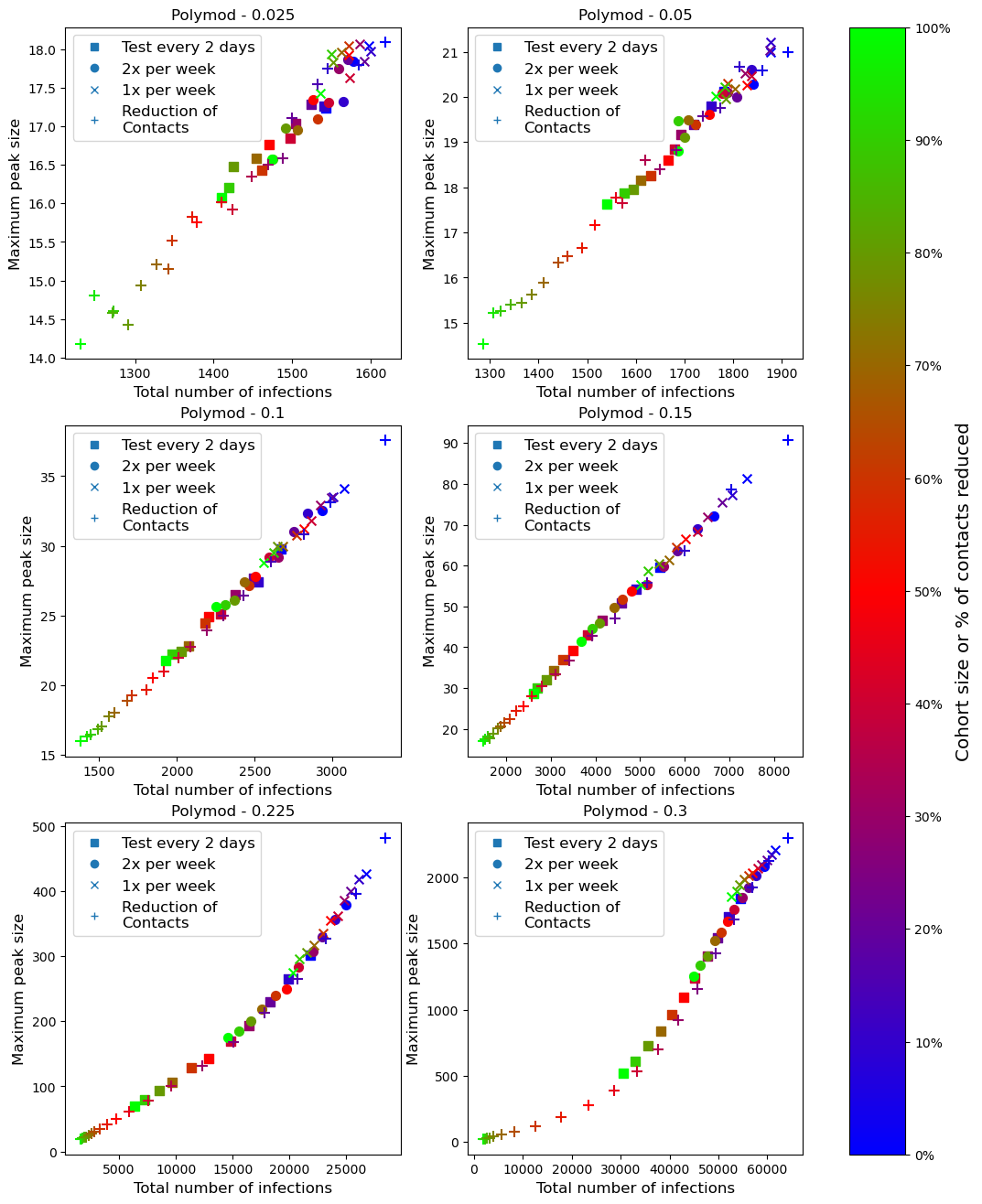}
        \caption{Contact patterns obtained from POLYMOD}
        \label{fig:asym_peak_cases_polymod}
    \end{subfigure}%
    \begin{subfigure}{.5\textwidth}
        \centering
        \includegraphics[width=\textwidth]{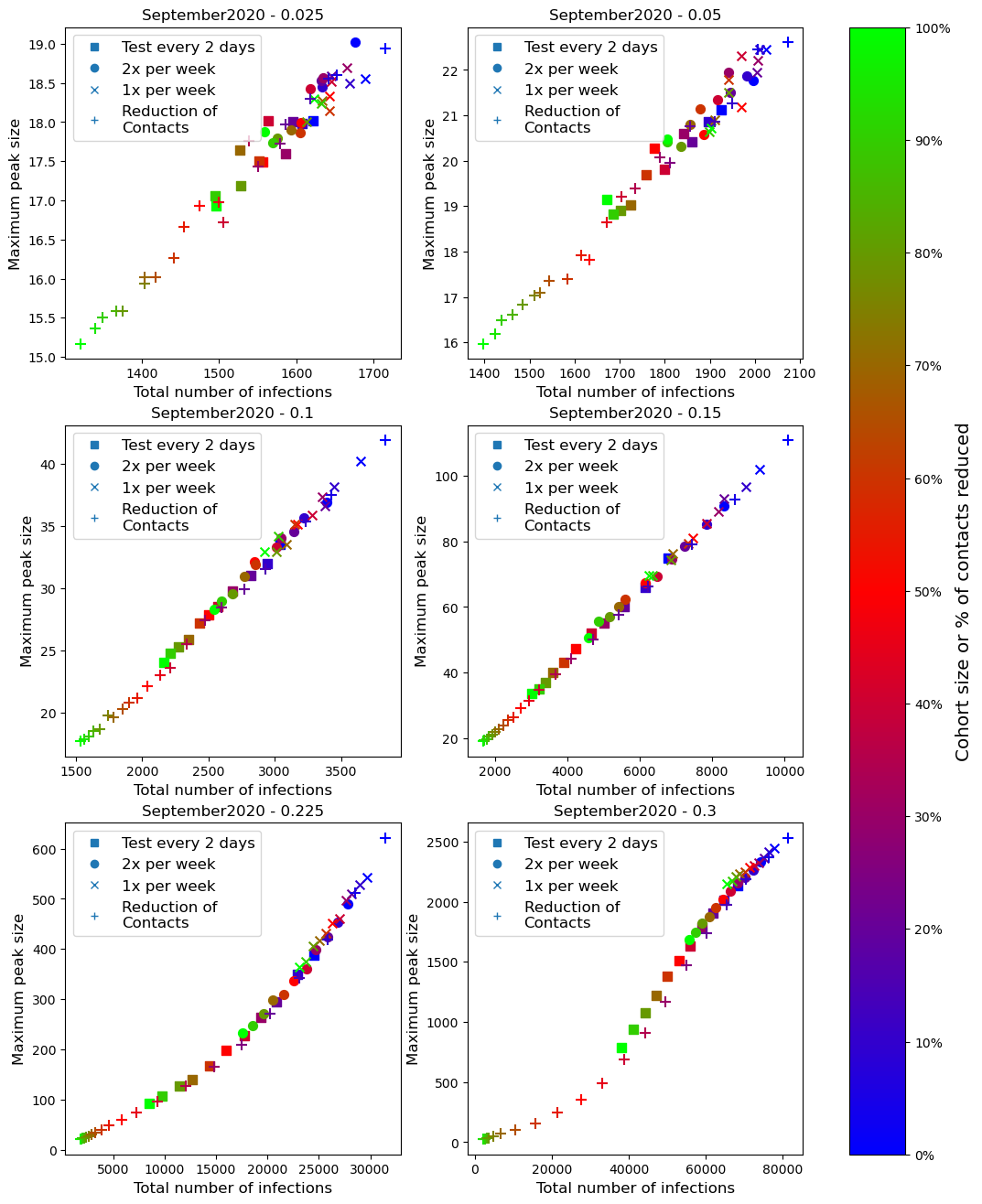}
        \caption{Contact patterns obtained from CoMix round 22 to 27}
        \label{fig:asym_peak_cases_comix}
    \end{subfigure}
    \caption{Comparison of testing strategies against reduction of contacts for different growth rates and contact patterns. The x-axis represents the number of total infections, averaged over 100 runs. The y-axis shows the maximum peak of new daily infections, averaged over 100 runs. The colour of each marker indicates the size of the regular tester cohort for the simulations with asymptomatic testing or the percentage of contacts removed from non-household layers for the simulations with contact reduction.}
    \label{fig:asym_peak_cases}
\end{figure*}

\clearpage

\begin{figure*}[!h]
    \centering
    \begin{subfigure}{.5\textwidth}
        \centering
        \includegraphics[width=0.95\textwidth]{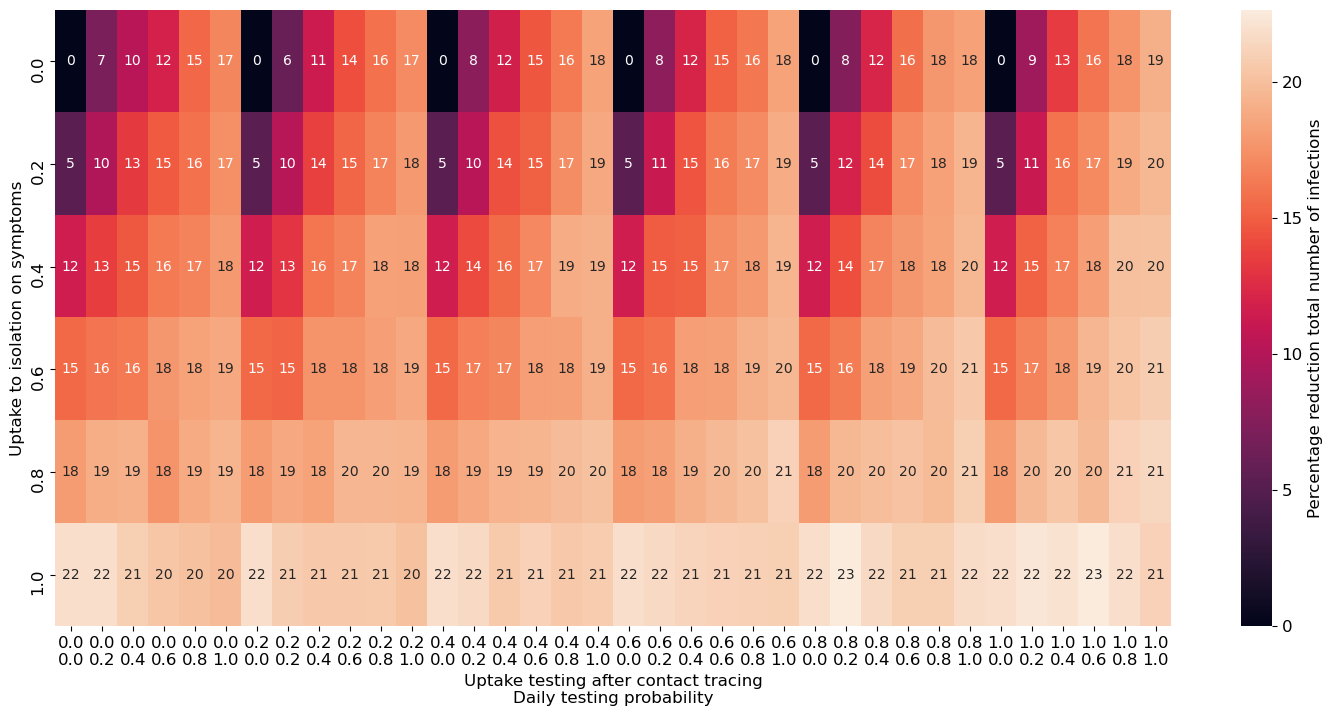}
        \caption{Growth rate = 0.05}
        \label{fig:symp_cases_0.05}
    \end{subfigure}%
    \begin{subfigure}{0.5\textwidth}
        \centering
        \includegraphics[width=0.95\textwidth]{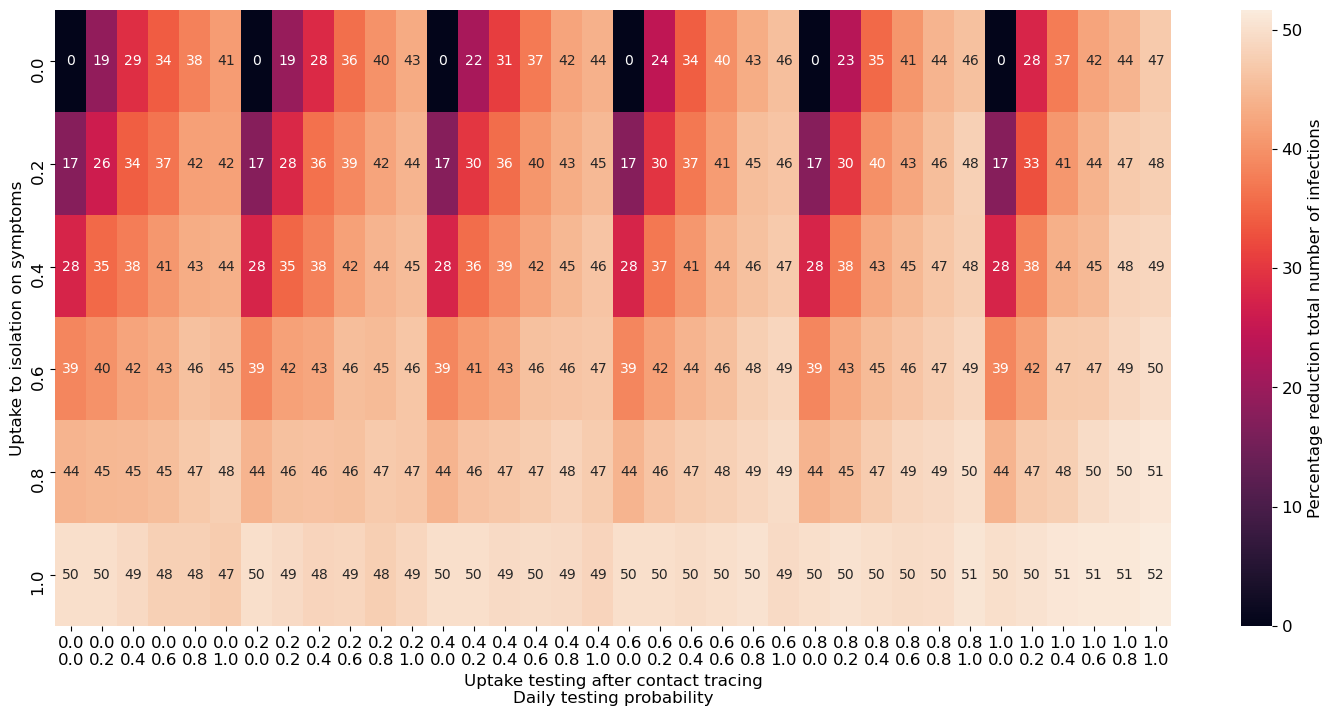}
        \caption{Growth rate = 0.1}
        \label{fig:symp_cases_0.1}
    \end{subfigure}
    \begin{subfigure}{.5\textwidth}
        \centering
        \includegraphics[width=0.95\textwidth]{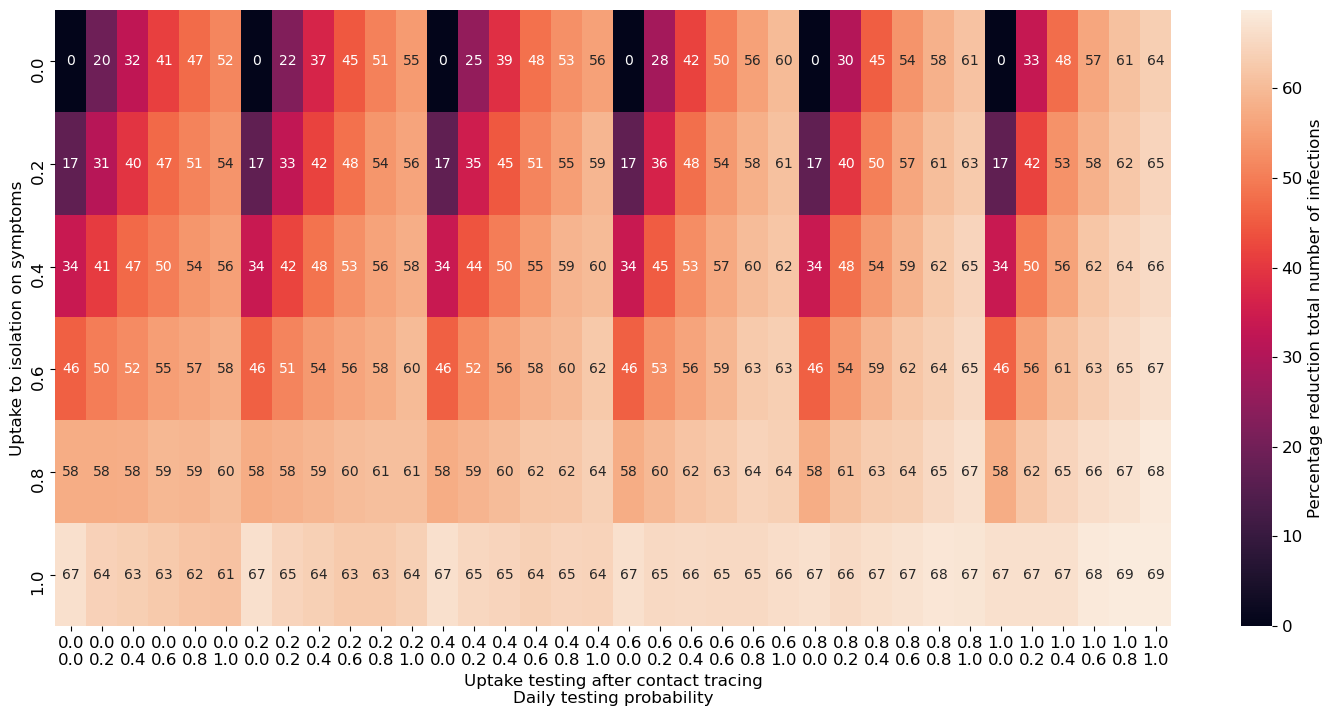}
        \caption{Growth rate = 0.15}
        \label{fig:symp_cases_0.15}
    \end{subfigure}%
    \begin{subfigure}{0.5\textwidth}
       \centering
        \includegraphics[width=0.95\textwidth]{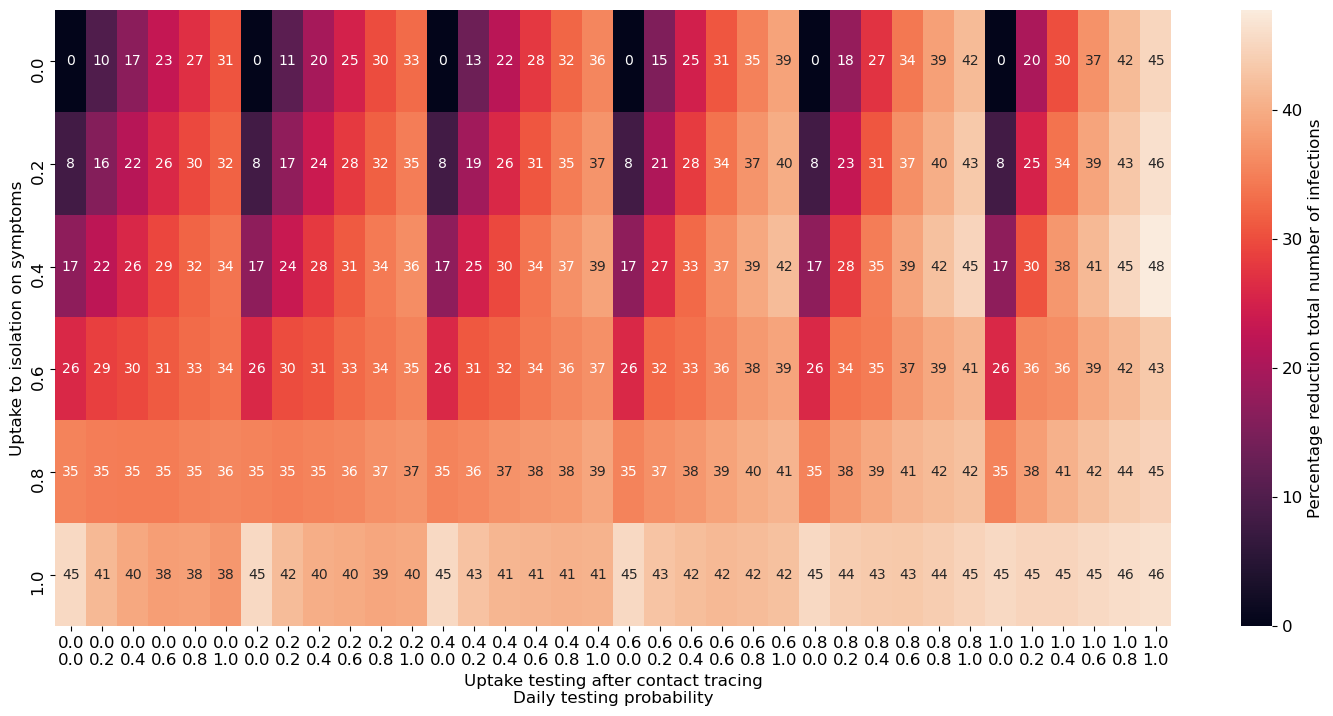}
        \caption{Growth rate = 0.225}
        \label{fig:symp_cases_0.225}
    \end{subfigure}
    \caption{Impact of increasing the uptake on isolation on symptoms, daily symptomatic testing and PCR testing after being traced as a contact of a positive case. Numbers shown are the percentage decrease in the total number of infections, compared to 0\% uptake on these interventions, for 0.05, 0.1, 0.15, 0.225 growth rates in POLYMOD contact pattern. The top line on the x axis represents the uptake to testing after contact tracing, the bottom line represents the daily probability to testing when symptomatic. The y axis represents uptake to isolation on COVID or ILI symptoms.}
    \label{fig:sym_cases_for_supp_material}
\end{figure*}

\clearpage

\begin{figure*}[!h]
    \centering
    \includegraphics[width=\textwidth]{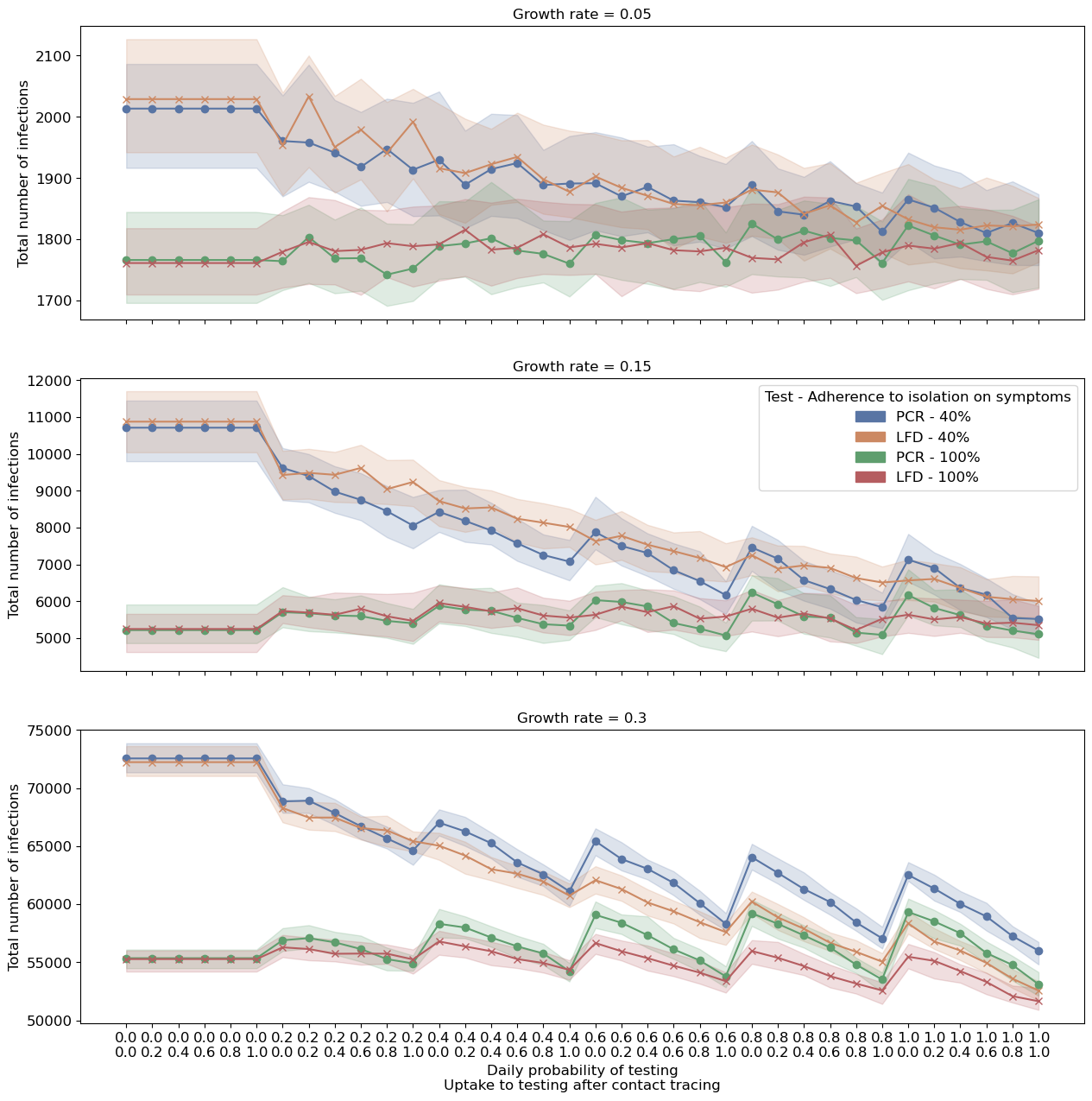}
    \caption{Comparing LFD and PCR tests and the impact of increasing the uptake of testing after being traced as a contact of a positive case and the daily probability of testing if symptomatic, for 3 different growth rates with contact patterns obtained from POLYMOD. The y-axis shows the total number of infections. The bottom line on the x axis represents the uptake to testing after contact tracing, the top line represents the daily probability to testing when symptomatic. Color indicates the type of test and the adherence to isolation on symptoms. Markers shown represent the median of 100 simulations and the shaded area represents the interquartile range.}
    \label{fig:sym_lfd_vs_pcr}
\end{figure*}

\begin{figure*}[!h]
        \centering
        \includegraphics[width=0.85\textwidth]{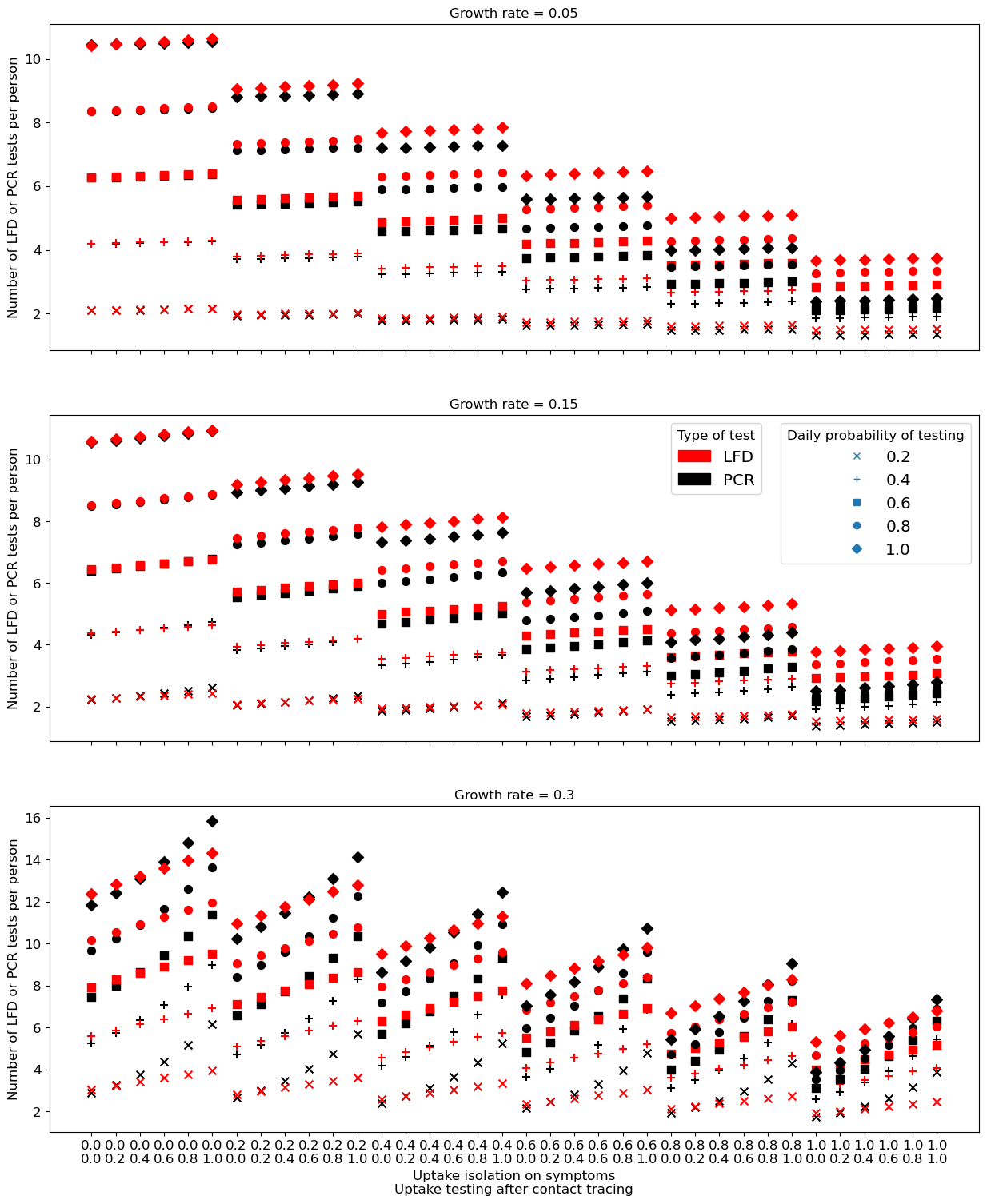}

    \caption{Comparison of number of tests required per person when using LFD or PCR tests for symptomatic or contact traced individuals, for 0.05, 0.15 and 0.3 growth rates, with contact patterns obtained from POLYMOD. The y-axis shows the mean number of tests per person, averaged over 100 runs. The bottom line on the x axis represents the uptake to testing after contact tracing, the top line represents the adherence to isolation on symptom onset.}
    \label{fig:lfd_vs_pcr_tests}
\end{figure*}

\clearpage

\begin{figure*}[!h]
    \centering
    \begin{subfigure}{0.5\textwidth}
        \centering
        \includegraphics[width=\textwidth]{asym_isolation_polymod}
        \caption{PCR as the test for symptomatic or contact traced individuals}
        \label{fig:asym_pcr_vs_lfd_isolation_pcr_polymod}
    \end{subfigure}%
    \begin{subfigure}{.5\textwidth}
        \centering
        \includegraphics[width=\textwidth]{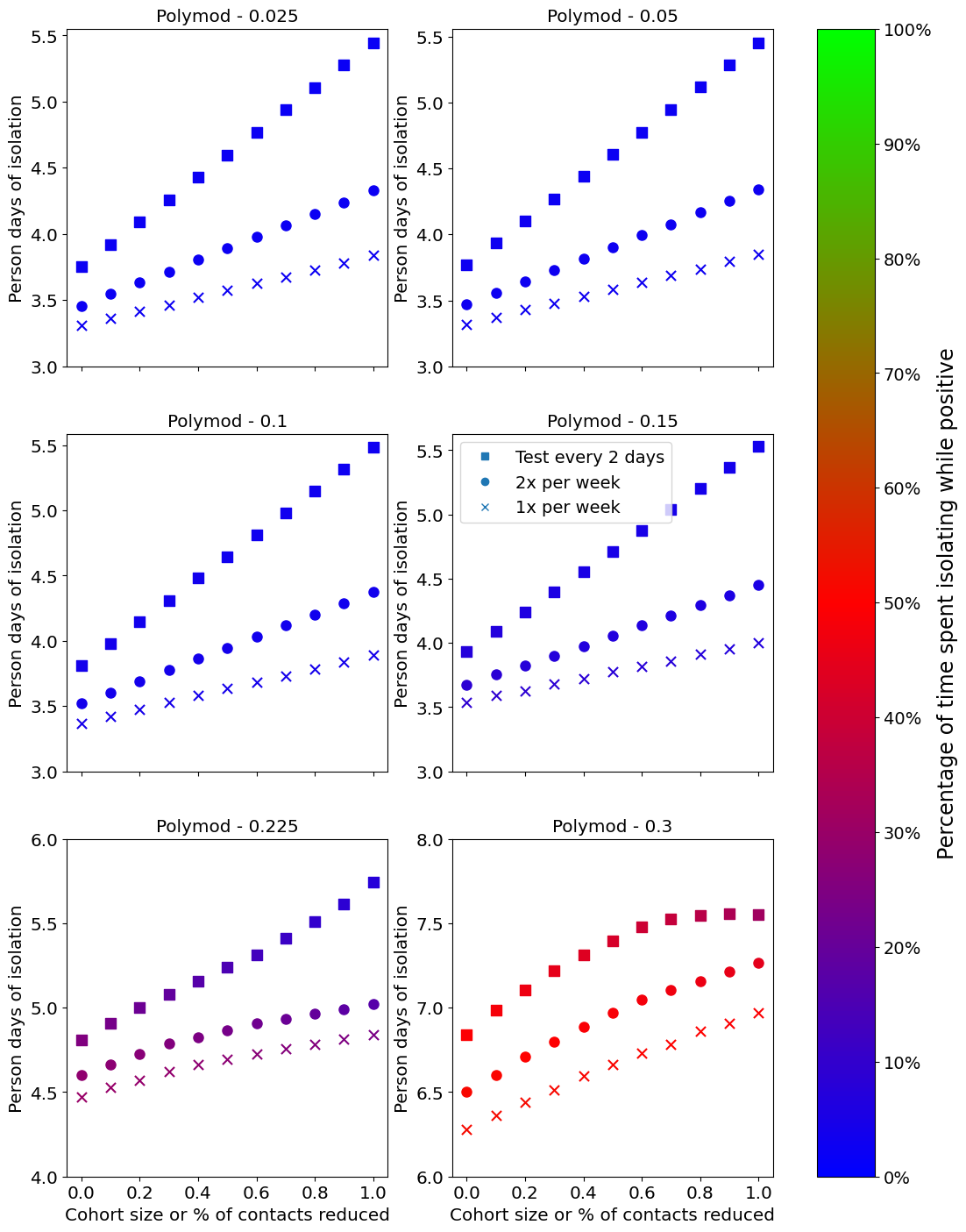}
        \caption{LFD as the test for symptomatic or contact traced individuals}
        \label{fig:asym_pcr_vs_lfd_isolation_lfd_polymod}
    \end{subfigure}

        \caption{Impact on person-days of isolation of increasing adherence to the different testing strategies, for different growth rates with contact patterns from POLYMOD, when using PCR (Figure~\ref{fig:asym_pcr_vs_lfd_isolation_pcr_polymod}) or LFD (Figure~\ref{fig:asym_pcr_vs_lfd_isolation_lfd_polymod}) as the test for symptomatic or contact traced individuals. The x-axis represents the size of the regular tester cohort for the simulations with asymptomatic testing or the percentage of contacts removed from non-household layers for the simulations with contact reduction. The y-axis shows the mean number of person days of isolation per population members over the 180 days of simulation, averaged over 100 runs. These include days in isolation waiting for test results (amongst true positives and true negatives); isolation days amongst those testing positive who were true positive; and isolation days among those testing positive who were false positives. The colour of each marker indicates the percentage of days spent in isolation while infected.}
        \label{fig:asym_pcr_vs_lfd_isolation}
\end{figure*}

\clearpage

\begin{figure*}[!h]
        \centering
        \includegraphics[width=0.90\textwidth]{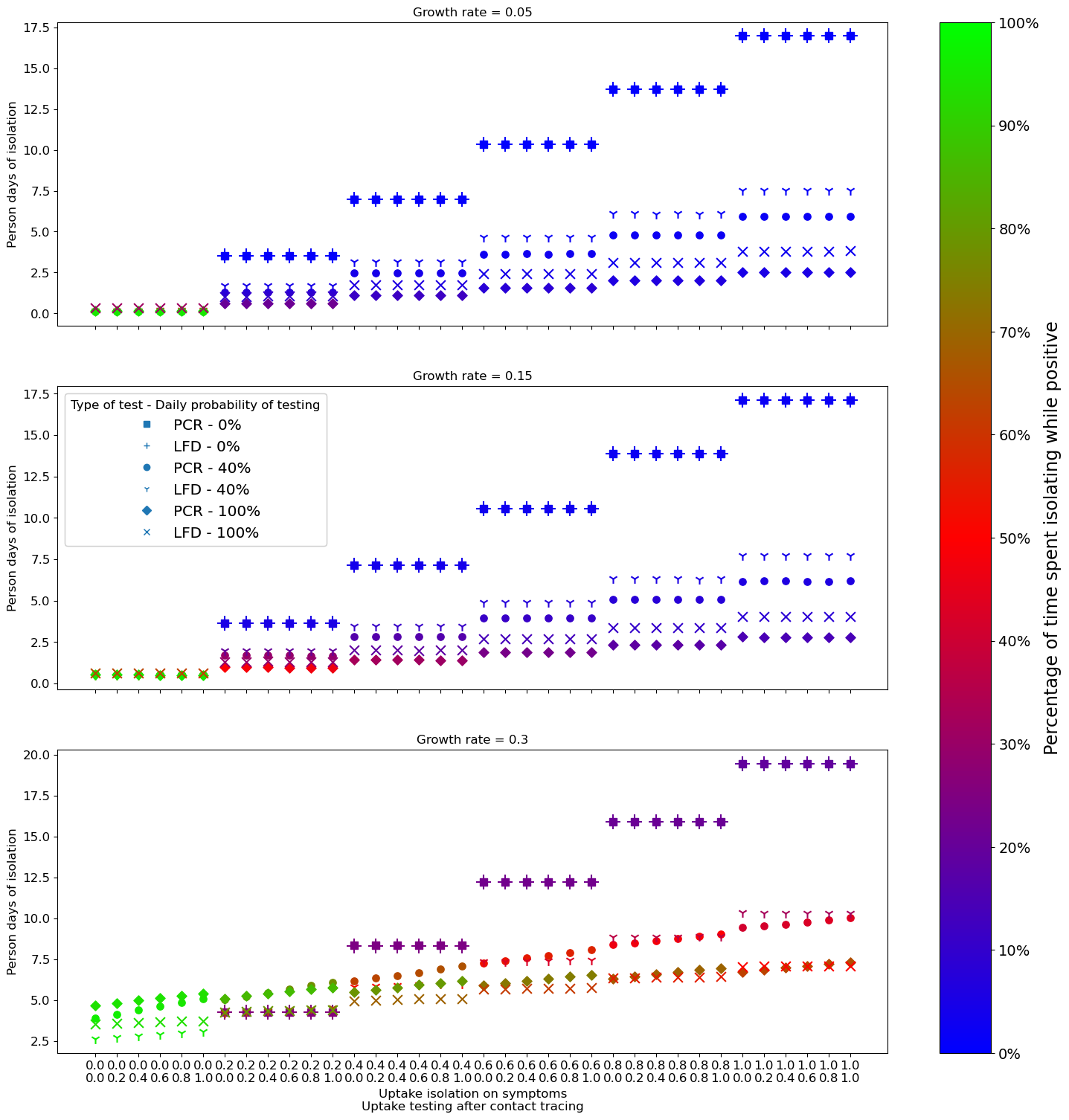}
        \caption{Impact on person-days of isolation of increasing adherence to isolation on symptoms, daily probability of testing and adherence to testing after contact tracing, for 0.05, 0.15 and 0.3 growth rates with contact patterns from POLYMOD, comparing the outcome when testing with PCR or LFD. The bottom line on the x axis represents the uptake to testing after contact tracing, the top line represents the adherence to isolation on symptom onset. The y-axis shows the mean number of person days of isolation per population members over the 180 days of simulation, averaged over 100 runs. These include days in isolation waiting for test results (amongst true positives and true negatives); isolation days amongst those testing positive who were true positive; and isolation days among those testing positive who were false positives. The colour of each marker indicates the percentage of days spent in isolation while infected.}
        \label{fig:sym_pcr_vs_lfd_isolation}
\end{figure*}

\end{document}